\documentclass[%
 reprint,
 amsmath,amssymb,
 pra,
floatfix,
]{revtex4-2}

\usepackage{graphicx}
\graphicspath{ {./Plots/Strain_Plots/}{./Plots/PSD_Plots/}{./Plots/Q-Scans/}{./Plots/Noise_Covariance/}{./Plots/Inference_Plots/}{./Plots/Fisher_Plots/} }

\usepackage{dcolumn}
\usepackage{bm}
\usepackage{hyperref}
\usepackage{amsmath} 
\usepackage{amssymb}
\usepackage{mathtools}
\usepackage{breqn}
\usepackage{cancel}
\usepackage{hyperref}
\usepackage{titlecaps}
\usepackage{subcaption}
\usepackage[colorinlistoftodos]{todonotes}
\usepackage{textcomp}

\usepackage{bm}

\usepackage{tikz}
\usetikzlibrary{calc}

\newcommand{\icg}{\affiliation{University of Portsmouth, Institute of Cosmology and Gravitation, Portsmouth PO1 3FX, United Kingdom}}

\begin{document}

\title{The Issues of Mismodelling Gravitational-Wave Data for Parameter Estimation}

\author{Oliver Edy}
 \email{oliver.edy@port.ac.uk}
 \icg
\author{Andrew Lundgren}%
 \email{andrew.lundgren@port.ac.uk}
 \icg
\author{Laura K. Nuttall}%
 \email{laura.nuttall@port.ac.uk}
 \icg

\date{\today}

\begin{abstract}
   Bayesian inference is used to extract unknown parameters from gravitational-wave signals. Detector noise is typically modelled as stationary, although data from the LIGO and Virgo detectors is not stationary. We demonstrate that the posterior of estimated waveform parameters is no longer valid under the assumption of stationarity. We show that while the posterior is unbiased, the errors will be under- or overestimated compared to the true posterior. A formalism was developed to measure the effect of the mismodelling, and found the effect of any form of non-stationarity has an effect on the results, but are not significant in certain circumstances. We demonstrate the effect of short-duration Gaussian noise bursts and persistent oscillatory modulation of the noise on binary-black-hole-like signals. In the case of short signals, non-stationarity in the data does not have a large effect on the parameter estimation, but the errors from non-stationary data containing signals lasting tens of seconds or longer will be several times worse than if the noise was stationary. Accounting for this limiting factor in parameter sensitivity could be very important for achieving accurate astronomical results. This methodology for handling the non-stationarity will also be invaluable for analysis of waveforms that last minutes or longer, such as those we expect to see with the Einstein Telescope.
\end{abstract}

\maketitle


\section{Introduction} \label{sec:introduction}

The Advanced LIGO detectors have been regularly detecting gravitational-wave signals since 2015 \cite{Aasi_2015}, joined by Advanced Virgo in 2017 \cite{Abbott_2017}, and the later addition of KAGRA in 2020 \cite{Akutsu_2020}. The signals observed so far have been produced by the coalescence of either a binary system of black holes (BBH) or neutron stars (BNS). The gravitational waves emitted contain information we can infer about the system, such as the mass of each component of the binary or their sky position. The signal is identified through techniques such as matched filtering \cite{Babak_2013}, which determines how closely the observed data matches with a theoretical waveform model known as a template. In practice, the output of the matched filter is calculated for a large bank of templates spanning the parameter space of expected signals \cite{Camp_2004}. However, the exact physical properties and physics of the system must be extracted using Bayesian analysis \cite{Veitch_2015}; these models evaluate a posterior probability density function of the unknown signal parameters that describe a specific model of the data. While Bayesian analysis is not designed to identify the template that fits the signal the best, when a suitable template has been found with matched filtering, Bayesian methods can obtain the parameter uncertainties more directly than the template search.

A gravitational-wave signal is fully described by 15 parameters, assuming negligible orbital eccentricity \cite{Abbott_2016_GW150914, Veitch_2015}; eight of these are intrinsic to the observed system, including the mass of each body before coalescence, or their spins, while the remaining seven depend on the observer (such as sky position, or the time $t_c$ in the observing frame at which the coalescence occurred). This means that even inference calculations of signals only a few seconds long are computationally expensive. Although techniques are being developed to improve processing time (for example, \cite{van_der_Sluys_2008, Zackay_2018, Smith_2016, Qi_2020, Morisaki_2020, Cornish_2021_RapidPE}), they are not yet mature. The most widely used inference codes for gravitational waves are LALInference \cite{Veitch_2015}, Bilby \cite{Ashton_2019, Romero_Shaw_2020}, and PyCBC Inference \cite{Biwer_2019}. These, as well as all other current methods, assume that the instrumental noise is both stationary and Gaussian (see, for example, \cite{Aasi_2013, Jaranowski_2005, Poisson_1995, Veitch_2015, Biwer_2019}). Note that we define stationarity to mean that the statistical properties of the noise do not change with time. Signal processing often uses wide-sense stationarity, where only the mean and covariance are assumed to be time-independent \cite{Perraudin_2017}. The fact that the statistical parameters of the noise (such as the spectrum shape or the intensity) do not vary with time means the data can be characterised \cite{Littenberg_2015}. Consequently, modelling and analysis of the noise in the frequency-domain can be simplified, which makes evaluation computationally efficient \cite{van_der_Sluys_2007}.

In reality, the gravitational-wave strain data from the LIGO and Virgo detectors is not stationary or Gaussian \cite{Blackburn_2008, Abbott_2016_TransientCharacterization, Abbott_2019, Abbott_2020, Chatziioannou_2019}. Hence, inference using the appropriate non-stationary model could not make use of the simpler form. There have been some methodologies proposed for handling the non-stationarity. For example, References \cite{Cornish_2020, Cornish_2021_BayesWave} proposed modelling the noise in a wavelet domain, where non-stationarity is more simply computed. This method has proven particularly useful in analysis of data containing glitches \cite{Chatziioannou_2021}. Other methods include estimating the variation in detector noise and using it to re-rank events to reduce the number of false events being reported \cite{Mozzon_2020}, or incorporating the uncertainty in the measurement of the power spectral density into parameter estimation \cite{Biscoveanu_2020}.

Typically, though, the assumption of stationarity and Gaussianity is taken instead \cite{Rover_2011}, simply as a matter of practicality because it enables a much simpler form for the likelihood to be used, where the `likelihood' refers to the model we choose for the data. Given that parameter estimation is known for its computational cost \cite{Veitch_2015}, a simpler model means that analysis is inexpensive. However, mismodelling the data in this way will affect the certainty with which we can estimate parameters from a gravitational-wave signal, a probability function known as the posterior. This could mean that uncertainties for parameters inferred from the system could be much smaller or larger than they really are. This compromise on the effectiveness of parameter estimation has already been explored in Reference \cite{Chatziioannou_2019}.

In this paper, we focus on the problems with non-stationary data and parameter estimation. While previous works such as those cited above have focused on handling the effects of non-stationarity, we concern ourselves with investigating how gravitational-wave analysis is affected if we do not account for the non-stationarity of the data. Specifically, our goal is to determine the effect that non-stationary noise will have on the estimated parameters, given that the assumed Bayesian model does not fit the data. This will establish the amount of non-stationarity which is allowable before the estimated parameters are significantly affected and it becomes necessary to explicitly account for the non-stationarity. In Section \ref{sec:motivation}, we first give a brief overview of how parameter estimation works, then consider whether approximating the data as stationary is a valid assumption to make. In Section \ref{sec:effect_of_nonst}, we suggest a model of Gaussian data with which we can determine deviations from stationarity by considering the covariance matrix. A few examples of how this might work are given in Section \ref{sec:applications}.

\section{Describing Non-Stationarity} \label{sec:motivation}

\subsection{Extracting Signals from Noisy Data} \label{ssec:likelihood_and_matched_filter}

A signal of known shape $h(t)$ is optimally extracted from noisy strain data $n(t)$ through a matched filter comparison with a number of templates \cite{Wainstein_1962}. The mass and spin of a signal must then be approximately the same as those that generated the matching template. Once a signal has been identified with techniques such as matched-filtering, analysis shifts to estimating the parameters of the compact-object binaries that generated the gravitational waves.

Methods for inferring details about a system are rooted in Bayes’ theorem, described in Ref \cite{Trotta_2008} as a generalised technique for understanding a system when working with incomplete information. Formally, for a set of parameters $\theta$, we represent this as
\begin{equation}
    p(\theta|d) = \frac{p(d|\theta) p(\theta)}{p(d)} \label{eq:Bayes},
\end{equation}
where we refer to $p(\theta|d)$ as the posterior of the data, $p(d|\theta)$ the likelihood, $p(d)$ the evidence, and $p(\theta)$ the prior. The posterior is the probability of obtaining the parameters $\theta$ given the data $d(t)$. The prior $p(\theta)$ describes an understanding of the system and the parameters $\theta$ before knowing anything about the data $d(t)$; the choice of prior is very important, as a poorly chosen prior will strongly bias the entire parameter estimation and likely lead to flat posteriors or wildly incorrect parameter estimations \cite{Robert_2010, Vitale_2017}.

The denominator $p(d)$ is called the evidence, and is a normalisation factor which can be used to directly compare one model of the data with another \cite{Gelman_1998}. Because of the way that Bayesian inference is implemented in determining the gravitational-wave parameters, it is normalisation-insensitive, and so it is rare to actually need to calculate the evidence \cite{Hogg_2018, Christensen_2001}. Note, though, that this is not always true, as the evidence is necessary in nested sampling, wherein new samples are drawn from a normalised prior probability (for example, Reference \cite{Speagle_2020} describes how this is implemented for the dynesty code).

The likelihood $p(d|\theta)$ is the probability of obtaining the data $d(t)$ observed, given the parameters $\theta$ being considered. In essence, it is the model we choose for the data, and the part we have the most control over.

As stated, the likelihood form $\mathcal{L}$ is assumed to be Gaussian of the form
\begin{equation}
    \mathcal{L} = e^{- \frac{1}{2} \langle n(f)^\dagger | n(f) \rangle } \label{eq:likelihood_matrix},
\end{equation}
where the square brackets denote the inner product averaged over all noise realisations. Note that the full derivation for this likelihood is described in Reference \cite{Wainstein_1962}.

With the assumption that the data is a multivariate Gaussian in the frequency domain, the noise covariance is diagonal, which we interpret to mean the frequencies are independent. In the case where the noise is one-dimensional, the Gaussian likelihood simplifies to
\begin{equation}
    \mathcal{L} = e^{- \frac{1}{2} n^\dagger(f) S^{-1} n(f)} \label{eq:likelihood_1D},
\end{equation}
where $S$ is the noise covariance.

\subsection{Validity of Stationarity Assumption} \label{ssec:validity_of_assumption}

We must consider whether it is indeed valid to model the data as stationary. First, we look at a description of stationary noise. According to Wold's theorem \cite{Scargle_2016}, any process $X$ can be represented as the sum of a deterministic part $D$ (which we shall ignore here) and a random part. The random part can then be further decomposed into a purely random component $R$ convolved with a purely deterministic process $C$. For a discrete dataset sampled at an evenly spaced set of times $n$, we have
\begin{equation}
    X_n = \sum_{n} C_{n-i} R_i + D_n = C * R + D \label{eq:Wold_theorem},
\end{equation}
where $R$ is completely uncorrelated, so that
\begin{equation}
    \langle R_i R_j \rangle = 0 \text{ if } i \neq j.
\end{equation}

This form suggests that stationary noise is simply white noise ($R$) being acted upon by a filter ($C$).

Although there is no simple classification of non-stationary noise \cite{Vajente_2020}, there are multiple different models (see, for example, \cite{Scargle_2016, Cramer_1961, Hebbal_2019, Kamen_1993, Loynes_1968, Huang_1998}). We concern ourselves with models that decompose non-stationary noise in a similar manner to stationary noise in Wold's theorem, this time into a family of non-deterministic stationary processes, or a family of deterministic non-stationary processes. We can compare this to the stationary case by now interpreting a time-varying filter to be acting on the white noise. The motivating example for this and follow-up work is non-stationary noise generated as the sum of some stationary Gaussian noise $n_1$ and a second Gaussian sample $n_2$ with oscillating amplitude $B(t)$:
\begin{equation}
    n(t) = A n_1(t) + B(t) n_2(t) \label{eq:non-stationary_form}.
\end{equation}
This example has been chosen for its simplicity, while also approximating the behaviour assumed by current LIGO calibration models: a stationary background noise $n_1$, and an unpredictable part $n_2(t)$ \cite{Sun_2020}. Similar models have already been incorporated into parameter estimation \cite{Vitale_2020, Payne_2020}, with strong agreement with the results of the LIGO data releases \cite{GWTC-1, LIGO_O2}. This indicates that as well as being a simple representation, the model is also a good fit for the real interferometer noise.

The assumption of stationarity is often fair to make providing that the amplitude $B(t)$ evolves slowly in time or is otherwise small. In the case of ground-based interferometers, the noise is measured to be nearly stationary at least over a period of several seconds – around the same duration as the signals that LIGO is able to detect \cite{GWTC-1, GWTC-2}.

However, Reference \cite{Littenberg_2015} has already shown these assumptions of stationarity break down for periods of $\sim$64 seconds, the timescales needed for the analysis of binary neutron star mergers. This was further explored in Reference \cite{Chatziioannou_2019}, which showed that parameter estimation will be noticeably compromised for segments of duration as small as 128 seconds.

Since we do not have a quantitative knowledge of the effect of non-stationarity, we cannot know how uncertain the posteriors might be even for short signals. Therefore, to understand any impact on parameter estimation, we present a formalism to describe how non-stationary data might differ from stationary data.

\subsection{Characterising the Noise} \label{ssec:noise_characterisation}

Gaussian noise can be completely characterised by its mean and covariance \cite{Chatziioannou_2019}, where the noise covariance represents random noise characteristics of the data. We define $n(f)$ as the discrete Fourier transform of a stretch of non-stationary noise and we represent it as a complex column vector $n$. Note that the noise is Gaussian \cite{Littenberg_2015}, so
\begin{equation}
    \langle n \rangle = 0,
\end{equation}
where the square brackets denote the ensemble average over all realisations of the noise.

We define the noise covariance $\Sigma$ as
\begin{equation}
    \Sigma = \langle n n^\dagger \rangle \label{eq:noise_covariance_non-stationary},
\end{equation}

and in the case of stationary noise, we instead express the same quantity as
\begin{equation}
    \Sigma_{\mathrm{stationary}} = S \label{eq:noise_covariance_stationary}.
\end{equation}

The covariance matrix reduces to a simpler form for stationary noise. In the frequency domain, the noise covariance $S$ becomes a positive diagonal matrix of the form
\begin{equation}
    S =
    \begin{pmatrix}
        S_1 & 0 & \ldots & 0 \\
        0 & S_2 & \ldots & 0 \\
        \vdots & \vdots & \ddots & \vdots \\
        0 & 0 & \ldots & S_N \\
    \end{pmatrix} \label{eq:diagonal_S},
\end{equation}
where $N$ is the number of data points. Since the likelihood is of the form $n^\dagger S^{-1} n$, we only care about values along the diagonal. This leaves a very simple likelihood form, as explored in Section \ref{ssec:derivation}.

When the noise is non-stationary, we have no robust method for estimating $\Sigma$ from the noise power spectrum, as we would for stationary noise. This is due to the near impossibility of characterising the data, due to the time variance of the statistical parameters \cite{Littenberg_2015}. As outlined in Reference \cite{Talbot_2020}, if we cannot assume stationarity, the method commonly used to estimate the power spectral density is to calculate an \emph{off-source} estimate, taking a mean of neighbouring segments, although this method will be more restrictive when considering the longer data segments needed to analyse a BNS.

Additional methods for calculating an estimate for the power spectral density have been proposed. For example, References \cite{Zackay_2019, Venumadhav_2019, Edwards_2020} discussed the concept of non-stationarities in detector data evolving the power spectral density (PSD) over time, referred to as PSD-drift. This causes a loss in sensitivity of the detectors, so correction methodology is offered by computing a running estimate of all matched filter overlaps. As Reference \cite{Huang_2020} noted, by correcting for PSD-drift, we see a notable difference in estimated parameters compared to the configurations assuming stationary data.

Even when the non-stationary covariance $\Sigma$ can be estimated, problems arise because it is not diagonal. This means the same simplification of the likelihood is not possible, and so the computation becomes far more complicated.

Using the likelihood given by Equation \eqref{eq:likelihood_1D}, and knowing the form of the noise covariance $S$, we will now derive a probe to investigate how greatly parameter estimation is affected by the non-stationarity of the data.

\section{Measuring the Effect of Non-Stationarity} \label{sec:effect_of_nonst}

\subsection{Deriving the Covariance Matrix of Waveform Parameters} \label{ssec:derivation}

We want to determine how greatly the estimated parameters of a particular waveform will be affected by the incorrect assumption of stationary noise. The best way to investigate this would be to derive the covariance matrix of the posterior distribution for the estimated waveform parameters, which describes how confidently we can measure each parameter.

The noise covariance is assumed to be a positive diagonal matrix $S$. The data, in the frequency domain, is written as a complex column matrix $d$. We use complex variables throughout the derivation, because the noise vector $n$ is complex in Fourier space. We then convert to a real matrix when we reach the final result (the covariance matrix of the waveform parameters).

We assume that the template is not a perfect fit for the signal. As in Reference \cite{Vallisneri_2008}, we approximate the template waveform $h$ as a linear series around the true signal $h_0$,
\begin{equation}
    h = h_0 + \omega \Vec{\theta} \label{eq:linear_waveform_approx},
\end{equation}
where the $\omega$ matrix is comprised of the derivatives of $h$ by each waveform parameter, and $\Vec{\theta}$ is the set of parameters (we will also encounter its Hermitian conjugate $\Vec{\theta}^\dagger$ in the derivation). Note that for simplicity we have only taken Equation \eqref{eq:linear_waveform_approx} to first-order.

We assume that the signal has been found in Gaussian data. The corresponding log-likelihood (up to a constant normalisation) is
\begin{dmath}
    \Lambda \propto -\frac{1}{2} \left( d - h|d - h \right) + \frac{1}{2} \left( d|d \right).
\end{dmath}

Note that the matrix notation of the noise-weighted inner product is given by
\begin{equation}
    \left( a | b \right) = 4 a^\dagger S^{-1} b,
\end{equation}
so the log-likelihood expression reduces to
\begin{equation}
    \Lambda \propto 2 \left( d^\dagger S^{-1} h + h^\dagger S^{-1} d - h^\dagger S^{-1} h \right) \label{eq:log-likelihood}.
\end{equation}

We decompose the strain data $d$ into the sum of the signal $h_0$ and the instrumental noise $n$ \cite{Chatziioannou_2019}:
\begin{equation}
    d = n + h_0  \label{eq:strain_components}.
\end{equation}

Using this substitution for $d$ in Equation \eqref{eq:log-likelihood}, as well as the expansion \eqref{eq:linear_waveform_approx}, the log-likelihood becomes
\begin{dmath}
    \Lambda \propto 2 \left( n^\dagger S^{-1} h_0 + h_0^\dagger S^{-1} n + h_0^\dagger S^{-1} h_0 + n^\dagger S^{-1} \omega \Vec{\theta} + \Vec{\theta}^\dagger \omega^\dagger S^{-1} n - \Vec{\theta}^\dagger \omega^\dagger S^{-1} \omega \Vec{\theta} \right).
\end{dmath}

Maximising $\Lambda$ with respect to $\Vec{\theta}$ and its Hermitian conjugate $\Vec{\theta}^\dagger$, we find the following expressions for the parameters at the maximum likelihood:
\begin{eqnarray}
    \Vec{\theta}_{\mathrm{ml}} &= \left( \omega^\dagger S^{-1} \omega \right)^{-1} \left( \omega^\dagger S^{-1} n \right), \nonumber \\
    \Vec{\theta}^\dagger_{\mathrm{ml}} &= \left( n^\dagger S^{-1} \omega \right) \left( \omega^\dagger S^{-1} \omega \right)^{-1} \label{eq:theta_mls}.
\end{eqnarray}

Note that we interpret $\Vec{\theta}_{\mathrm{ml}}$ as complex to simplify calculations, and hence the presence of $\Vec{\theta}^\dagger_{\mathrm{ml}}$. We convert $\Vec{\theta}$ into a fully real matrix in Section \ref{ssec:nuisance_parameters}.

From References \cite{Vallisneri_2011, Sathyaprakash_2009}, we note that
\begin{equation}
    \langle \Vec{\theta}_{\mathrm{ml}} \rangle = 0, \label{eq:theta_av}
\end{equation}
so long as the template matches the true signal, which shows that the posterior is still centered around the same point. Therefore, to linear order at least, mismodelling the data will only deform the shape of the posterior (thereby over- or under-representing the confidence contours), but does not cause any bias to the estimated parameters.

For ease of reading, the $\mathrm{ml}$ subscript will be dropped from $\Vec{\theta}$ for the rest of the text unless otherwise stated.

By multiplying the two quantities in Equation \eqref{eq:theta_mls} together, we create the covariance matrix for $\Vec{\theta}$:
\begin{equation}
    \langle \Vec{\theta}\Vec{\theta}^\dagger \rangle = \left( \omega^\dagger S^{-1} \omega \right)^{-1} \left( \omega^\dagger S^{-1} \langle n n^\dagger \rangle S^{-1} \omega \right) \left( \omega^\dagger S^{-1} \omega \right)^{-1}.
\end{equation}

As the noise covariance is defined in Equation \eqref{eq:noise_covariance_non-stationary}, we see that the covariance matrix becomes
\begin{equation}
    \langle \Vec{\theta}\Vec{\theta}^\dagger \rangle = \left( \omega^\dagger S^{-1} \omega \right)^{-1} \left( \omega^\dagger S^{-1} \Sigma S^{-1} \omega \right) \left( \omega^\dagger S^{-1} \omega \right)^{-1} \label{eq:covariance_matrix_non-stationary}
\end{equation}
for non-stationary noise. However, in the case when $S$ matches the true covariance $\Sigma$, then Equation \eqref{eq:covariance_matrix_non-stationary} greatly simplifies. This is the case for stationary noise, because the quantity $\langle n n^\dagger \rangle = S$ when averaged over infinite noise realisations, as stated in Equation \eqref{eq:noise_covariance_stationary}, resulting in the familiar
\begin{equation}
    \langle \Vec{\theta} \Vec{\theta}^\dagger \rangle = \left( \omega^\dagger S^{-1} \omega \right)^{-1} \label{eq:covariance_matrix}.
\end{equation}

This result is comparable to the inverse of the Fisher matrix \cite{Coe_2009}. We intend to use deviations from the stationary covariance matrix (Equation \eqref{eq:covariance_matrix}) to show how mismodelling non-stationary data can affect parameter estimation.

\subsection{Distinction Between \texorpdfstring{$\Vec{\theta}$}{o} and \texorpdfstring{$\Vec{\theta}_{\mathrm{ml}}$}{o}} \label{ssec:theta_ml}

To fully understand the difference between $\Vec{\theta}$ and $\Vec{\theta}_{\mathrm{ml}}$, we must consider the basis of parameter estimation in Bayes' theorem \cite{Trotta_2008}, as given by Equation \eqref{eq:Bayes}.

$\langle \Vec{\theta}_{\mathrm{ml}} \Vec{\theta}^\dagger_{\mathrm{ml}} \rangle$ is the covariance matrix of the parameter set $\Vec{\theta}$ at the maximum likelihood, averaged over all noise realisations. This makes $\Vec{\theta}^\dagger_{\mathrm{ml}}$ a point estimate, not a random variable; we have sought point values for $\Vec{\theta}$ which maximise the posterior $p(\Vec{\theta}|d)$, and treat the term $\frac{p(\Vec{\theta})}{p(d)}$ as a constant. In this way, we do not inject any prior beliefs about the parameters into the calculation for likely values of $\Vec{\theta}$, and so effectively ignore the prior.

A full Bayesian treatment would estimate the entire distribution for $p(\Vec{\theta}|d)$, treating $\Vec{\theta}$ as an array of random variables. Although the inclusion of the prior would more tightly constrain the covariance matrix, we consider only the covariance of the maximum likelihood here both because its inverse is the proper definition of the Fisher matrix \cite{Vallisneri_2011}, and because the prior does not change considerably in the regime of high signal-to-noise ratio (SNR) \cite{Sathyaprakash_2009}. This last point in particular is important as we are concerning ourselves with whether analysis of a real signal is affected by the non-stationarity of the data.

\subsection{Dealing with Nuisance Parameters}  \label{ssec:nuisance_parameters}

To minimise dimensionality during calculations, the covariance matrix has been treated as complex, but the parameters of $\Vec{\theta}$ are real. It is at this point that we decompose the complex matrix from Equation \eqref{eq:covariance_matrix} into a fully real matrix, with each column or row representing a parameter. Since the matrix has been treated as complex, however, there are several rows and columns that now contain nuisance parameters, which are the imaginary counterparts to parameters which do exist in the waveform. For example, we show in Section \ref{ssec:toy_model} by taking the derivative of the waveform by the phase $\phi$, that it is simply the complex counterpart to the derivative by the amplitude $\mathcal{A}$. Similarly, the other parameters of $h(f)$ have complex counterparts, although these are non-physical, and not represented in the waveform $h(f)$.

Given that Equation \eqref{eq:covariance_matrix_non-stationary} is a covariance matrix, it is possible to \emph{fix} these nuisance parameters. This is done by inverting it into a Fisher matrix, removing the rows and columns corresponding to these nuisance parameters, and inverting back to find the new covariance matrix. Therefore, we are left with a covariance matrix for only the physical parameters.

For now, let us only consider the parameters $\mathcal{A}$, $\phi$, and $t_c$. Given we have decomposed the matrix to be fully real, we would also need to consider the complex counterpart to the parameters. For $\phi$, that is simply $\mathcal{A}$, as mentioned above, but for $t_c$, we would need to introduce an additional nuisance parameter, $\xi$, which will shortly be fixed from the covariance matrix.

The log-likelihood would be of a form similar to
\begin{widetext}
\begin{dmath}
    \Lambda = \Vec{\theta}^T C^{-1} \Vec{\theta} =
    \begin{pmatrix}
        \Delta \mathcal{A} &
        \Delta \phi &
        \Delta \xi &
        \Delta t_c
    \end{pmatrix}
    \begin{pmatrix}
        \sigma_{\mathcal{A}\mathcal{A}} & \sigma_{\mathcal{A}\phi} & \sigma_{\mathcal{A}\xi} & \sigma_{\mathcal{A}t_c} \\
        \sigma_{\phi\mathcal{A}} & \sigma_{\phi\phi} & \sigma_{\phi\xi} & \sigma_{\phi t_c} \\
        \sigma_{\xi\mathcal{A}} & \sigma_{\xi\phi} & \sigma_{\xi\xi} & \sigma_{\xi t_c} \\
        \sigma_{t_c\mathcal{A}} & \sigma_{t_c\phi} & \sigma_{t_c\xi} & \sigma_{t_ct_c}
    \end{pmatrix}
    \begin{pmatrix}
        \Delta \mathcal{A} \\
        \Delta \phi \\
        \Delta \xi \\
        \Delta t_c
    \end{pmatrix} \label{eq:fixing_parameter_example},
\end{dmath}
\end{widetext}
where $C$ is our covariance matrix.

We are not actually interested in the nuisance parameter $\xi$, as it is not a term of the waveform $h(f)$. This means we know that $\xi = 0$ in the waveform, and it can be removed from the Fisher matrix by fixing it. This involves stripping out any row or column that would be multiplied by $\xi$. Therefore, Equation \eqref{eq:fixing_parameter_example} reduces to
\begin{equation}
    \Lambda =
    \begin{pmatrix}
        \Delta \mathcal{A} &
        \Delta \phi &
        \Delta t_c
    \end{pmatrix}
    \begin{pmatrix}
        \sigma_{\mathcal{A}\mathcal{A}} & \sigma_{\mathcal{A}\phi} & \sigma_{\mathcal{A}t_c} \\
        \sigma_{\phi\mathcal{A}} & \sigma_{\phi\phi} & \sigma_{\phi t_c} \\
        \sigma_{t_c\mathcal{A}} & \sigma_{t_c\phi} & \sigma_{t_c t_c}
    \end{pmatrix}
    \begin{pmatrix}
        \Delta \mathcal{A} \\
        \Delta \phi \\
        \Delta t_c
    \end{pmatrix}.
\end{equation}

The Fisher matrix used here can then be inverted back to get the new covariance matrix.

There are other parameters that we want to remove because they are not useful in the analysis, such as the phase $\phi$, which tells us no extra information than the amplitude $\mathcal{A}$. These cannot be treated in the same way because they are real parameters of $h(f)$ and we do not know their value to be able to fix them. Instead, we must marginalise over these parameters, integrating them out of the probability function. An example of such is given in Section \ref{ssec:toy_model}.

\section{Applications}  \label{sec:applications}

\subsection{2D Toy Model} \label{ssec:toy_model}

We propose a toy model to demonstrate the calculation and subsequent handling of the covariance matrix of a waveform's parameters. Let us now consider the simplest case, for which we assume that we know all the parameters of the waveform $h(f)$, except for the amplitude $\mathcal{A}$, and the phase $\phi$. The waveform can now be described as follows:
\begin{equation}
    h (f) = h_0 (f) e^{\mathcal{A}+i\phi},
\end{equation}
where $h_0$ is the waveform of known parameters.

The corresponding derivatives of $h(f)$ by these two parameters are
\begin{align}
    \partial_{\mathcal{A}}h &= h, & \partial_{\phi}h &= ih.
\end{align}

Taking the linear waveform approximation as in Equation \eqref{eq:linear_waveform_approx}, and recalling that $\omega$ is the derivative of the waveform with the parameters, we find an $N\times1$ row matrix of
\begin{align}
    \omega
    &=
    \begin{pmatrix}
        \partial_{\Vec{\theta}}h
    \end{pmatrix}\\
    &=
    \begin{pmatrix}
        \partial_{\mathcal{A}}h + \partial_{\phi}h
    \end{pmatrix}\\
    &=
    \begin{pmatrix}
        h + i h
    \end{pmatrix},
\end{align}
where $N$ is the length of $h$.

Using this $\omega$ in Equation \eqref{eq:covariance_matrix}, we re-express the covariance matrix as
\begin{align}
    \langle \Vec{\theta} \Vec{\theta}^\dagger \rangle &= \left(
    \begin{pmatrix}
        h^\dagger - i h^\dagger
    \end{pmatrix}
    S^{-1}
    \begin{pmatrix}
        h + i h
    \end{pmatrix}
    \right)^{-1} \nonumber \\
    &=
    \begin{pmatrix}
        2 h^\dagger S^{-1} h + i \left( \cancel{h^\dagger S^{-1} h} - \cancel{h^\dagger S^{-1} h} \right)
    \end{pmatrix}^{-1} \nonumber \\
    &=
    \begin{pmatrix}
        \sigma_1 + i \sigma_2
    \end{pmatrix},
\end{align}
where $\sigma_1 = 2 h^\dagger S^{-1} h$ and $\sigma_2 = 0$.

The covariance matrix can be decomposed into a fully real $2\times2$ matrix by separating out the real and imaginary values as follows:
\begin{equation}
    \langle \Vec{\theta} \Vec{\theta}^\dagger \rangle =
    \begin{pmatrix}
        \sigma_1 & \sigma_2 \\
        - \sigma_2 & \sigma_1
    \end{pmatrix}.
\end{equation}
With the matrix decomposition demonstrated, from here on $\sigma_2$ will be explicitly referred to as 0.

The corresponding log-likelihood is
\begin{equation}
    \Lambda
    =
    \begin{pmatrix}
        \Delta \mathcal{A} & \Delta \phi
    \end{pmatrix}
    \begin{pmatrix}
        \sigma_1 & 0 \\
        0 & \sigma_1
    \end{pmatrix}
    \begin{pmatrix}
        \Delta \mathcal{A} \\
        \Delta \phi
    \end{pmatrix} \label{eq:log-likelihood_toymodel}.
\end{equation}

Note the form of the covariance matrix:
\begin{equation}
    \centering
    \begin{tabular}{ r | *{2}{c} }
    \enskip & $\Delta \mathcal{A}$ & $\Delta \phi$ \\ \hline
    $\Delta \mathcal{A}$ & $\sigma_1$ & 0 \\
    $\Delta \phi$ & 0 & $\sigma_1$ \\
    \end{tabular}
    \label{tab:Covariance_ToyModel}
\end{equation}

The covariance between the amplitude term $\Delta \mathcal{A}$ and its complex counterpart $\Delta \phi$ is 0, while covariance for $\Delta \mathcal{A} \Delta \mathcal{A}$ and $\Delta \phi \Delta \phi$ is the same (here, $\sigma_1$). This is true for any parameters and their complex counterparts.

We can marginalise over any parameter we are not interested in to remove it from the likelihood, and further simplify the likelihood form for computational efficiency.

We receive the same information from both the amplitude and the phase, so we do not need both. For demonstration purposes, we are not interested in the phase of the gravitational wave, and so marginalise over the parameter to remove it from the likelihood in Equation \eqref{eq:log-likelihood_toymodel}. For a Gaussian probability, marginalising over a parameter produces an identical result to maximising by that parameter (up to normalisation), so for ease, we maximise by $\phi$ instead,
\begin{equation}
    \frac{\partial \Lambda}{\partial \Delta \phi}
    = \frac{\partial}{\partial \Delta \phi} \left( \sigma_1 \Delta \mathcal{A}^2 + \sigma_1 \Delta \phi^2 \right) = 0.
\end{equation}

We can see that $\Delta \phi = 0$, and so the simplified form of the log-likelihood is

\begin{equation}
    \Lambda
    =
    \begin{pmatrix}
        \Delta \mathcal{A}
    \end{pmatrix}
    \begin{pmatrix}
        \sigma_1
    \end{pmatrix}
    \begin{pmatrix}
        \Delta \mathcal{A}
    \end{pmatrix}.
\end{equation}

Interestingly, in this simple case, this is the same result as if we had fixed $\phi$, but typically fixing a parameter and marginalising over it would produce very different matrices. Otherwise, this toy model provides a simple demonstration of how we can find and then simplify the covariance matrix for a system. In the next section, we then show how the covariance matrix might be affected by non-stationarity in a real-life scenario.

\subsection{Visualising the Effect of Non-Stationarity} \label{ssec:example_of_effect_of_nonst}

Although we have described the effect of non-stationarity in a theoretical capacity, we have yet to visualise it. To demonstrate, we shall compare the idealised stationary Gaussian model against two different non-stationary models which we name Models A and B. Model A localises the non-stationary noise over a short period of time, whereas the non-stationarity in Model B is present over the entire dataset, and hence models an extended period of non-stationarity.

For Model A, we have generated Gaussian data as described by Equation \eqref{eq:non-stationary_form}, with the time-varying amplitude $B(t)$ set to 0 at all times except for a period of 8 seconds centred around a time $t_0$:
\begin{equation}
    B(t) = 
    \begin{cases}
        \textrm{Tukey}(\alpha=0.5) & t_0 - 4 < t_0 < t_0 + 4 \\
        0 &\textrm{otherwise}.
    \end{cases} \label{eq:Tukey}
\end{equation}

In the latter instance, the amplitude will describe a Tukey window with the shape parameter $\alpha = 0.5$. The window was specifically chosen to last 8 seconds because it exceeds the length of the majority of signals measured by LIGO to date. Therefore, we will be able to determine whether the non-stationarity would have an effect on signals of comparable length.

In the case of Model B, the amplitude $B(t)$ described in Equation \eqref{eq:non-stationary_form} is sinusoidal, oscillating at 3.2 Hz.

These models were chosen as simplified approximations to real phenomena seen in the LIGO interferometers. A short period of stationarity as seen in Model A is comparable to a short burst of non-stationarity that is created by environmental events such as thunderstorms. Model B is instead an approximation for extended periods of non-stationarity, analogous to microseismic ground-motion \cite{Biswas_2020}. The realistic counterparts are never expected to become as extreme as the models described here, which were chosen to make the effect of non-stationarity on parameter estimation suitably explicit. Additionally, due to the very different forms of noise that we have created, it is not appropriate to compare the results of Model A or B with the other, but only against the stationary Gaussian model.

These three data samples (the stationary Gaussian noise, and Models A and B) will be 40 seconds long. This is considered a suitable duration to produce reliable results, and not too long a duration to not be relevant to the shorter data segments analysed by inference codes.

\begin{figure}
    \centering
    \begin{subfigure}[b]{0.475\textwidth}
        \includegraphics[width=\textwidth]{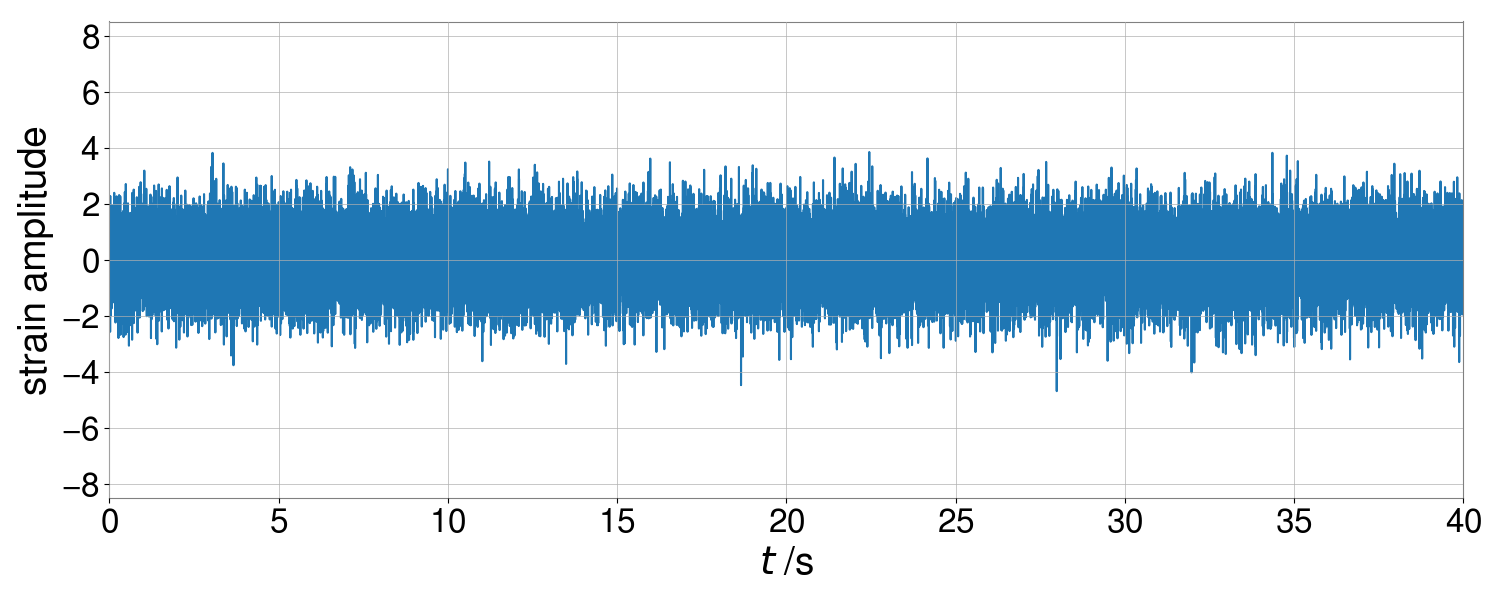}
        \caption{Stationary Gaussian noise.}
        \label{fig:Strain_Gaussian}
    \end{subfigure}
    \begin{subfigure}[b]{0.475\textwidth}
        \includegraphics[width=\textwidth]{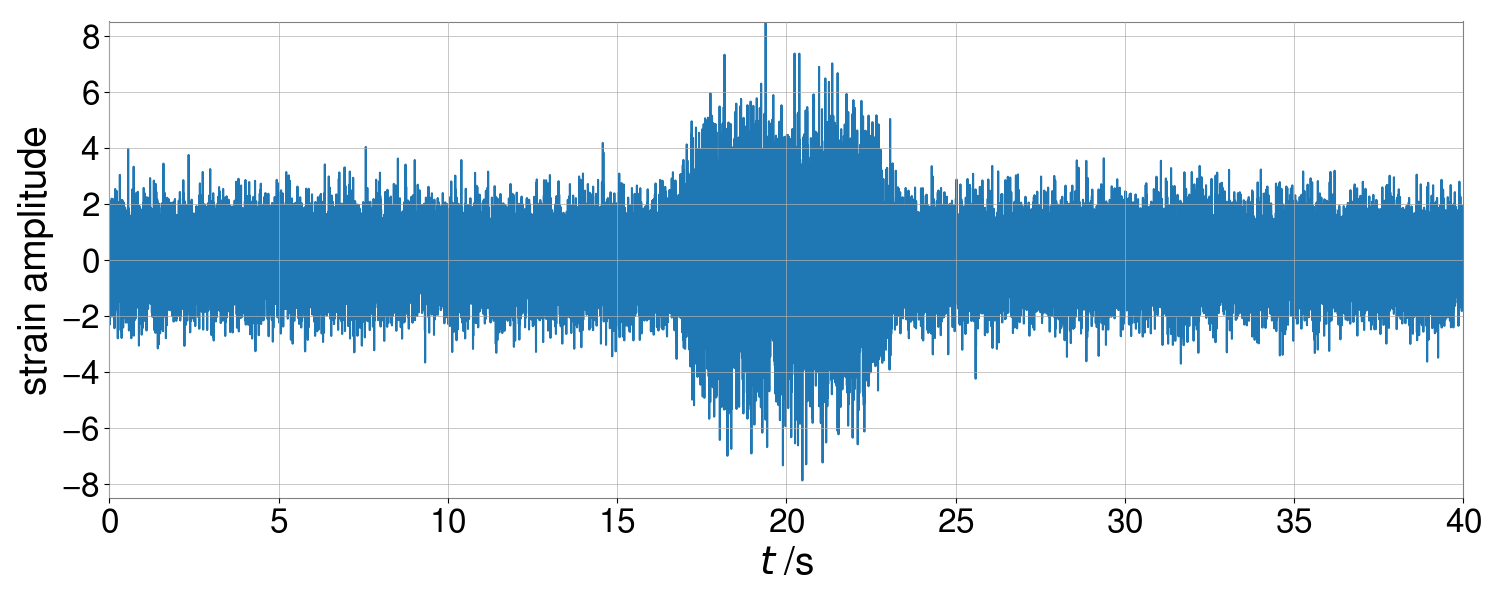}
        \caption{Gaussian noise with an 8-second Tukey window applied at 16 to 24 seconds.}
        \label{fig:Strain_GaussianModulation}
    \end{subfigure}
    \begin{subfigure}[b]{0.475\textwidth}
        \includegraphics[width=\textwidth]{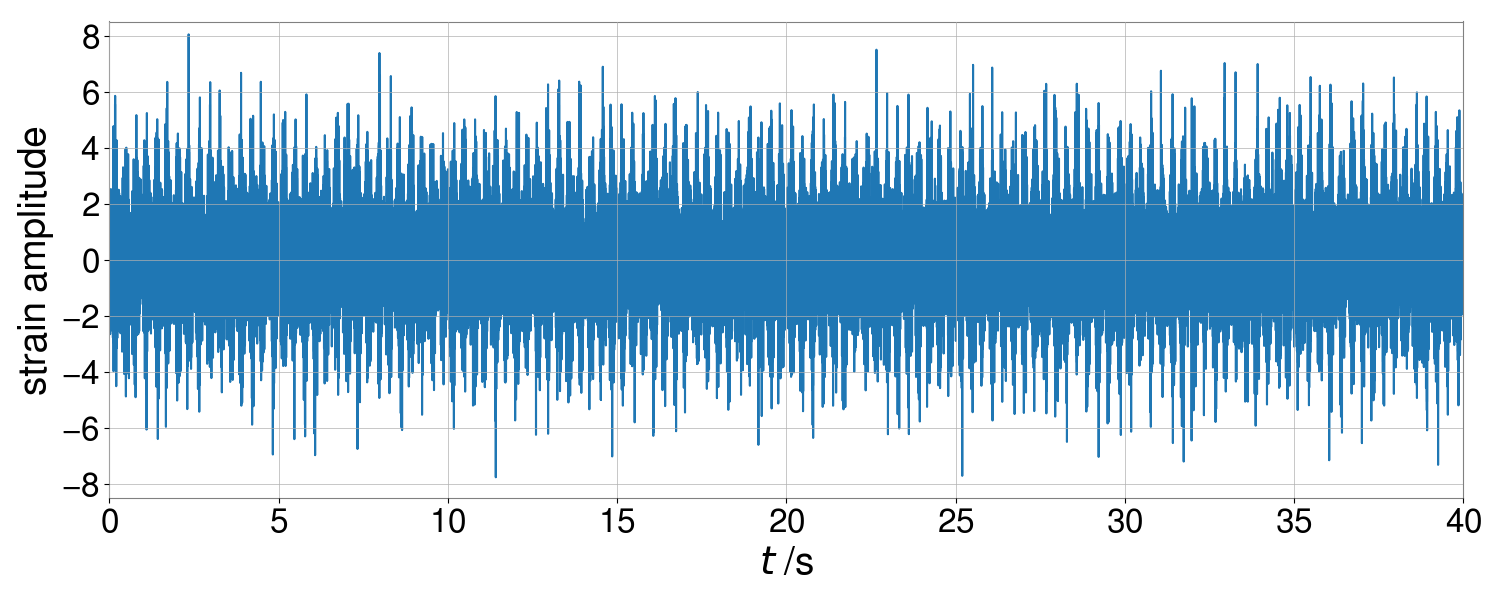}
        \caption{Gaussian noise with an oscillating amplitude of 3.2 Hz.}
        \label{fig:Strain_GaussianTukey}
    \end{subfigure}
    \caption{Strain plots for three artificially generated timeseries over the entire 40-second sample.}
    \label{fig:Strain_Comparison}
\end{figure}

\begin{figure}
    \centering
    \begin{subfigure}[b]{0.475\textwidth}
        \includegraphics[width=\textwidth]{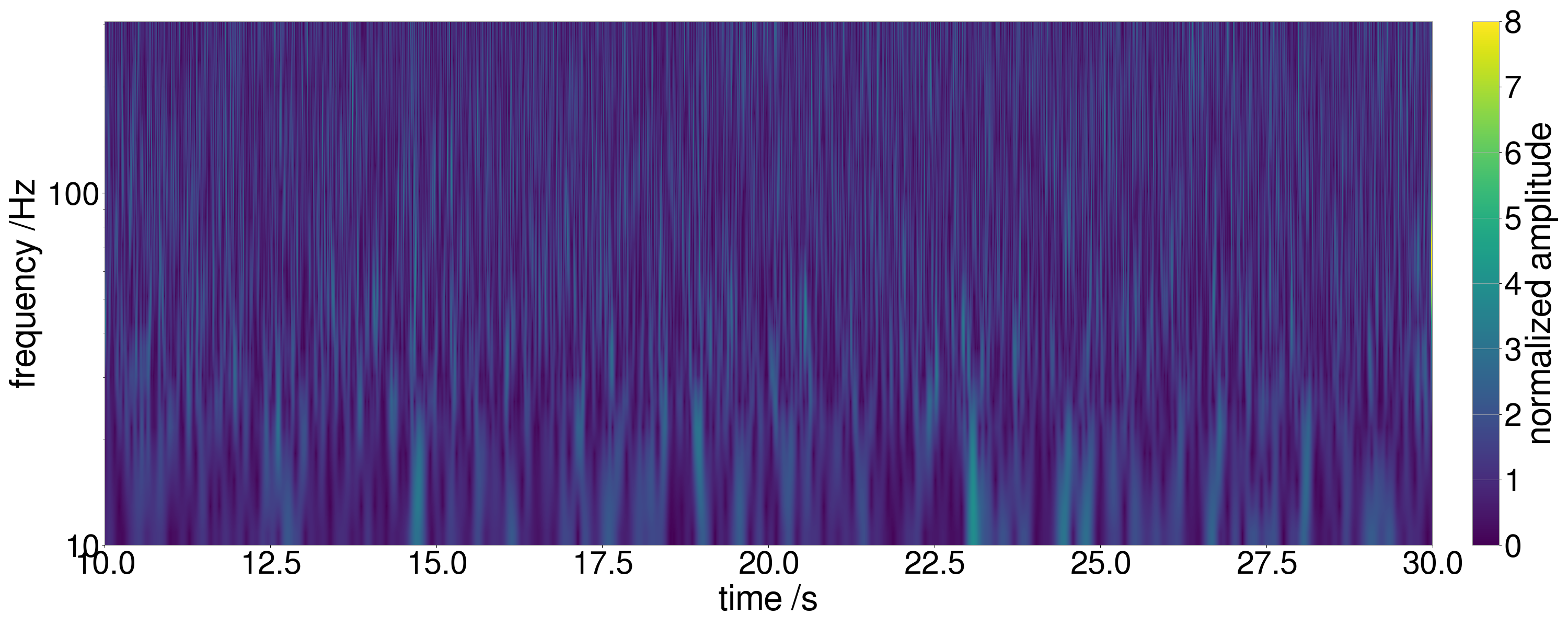}
        \caption{Time-frequency spectrogram of stationary Gaussian noise.}
        \label{fig:QScan_Gaussian}
    \end{subfigure}
    \begin{subfigure}[b]{0.475\textwidth}
        \includegraphics[width=\textwidth]{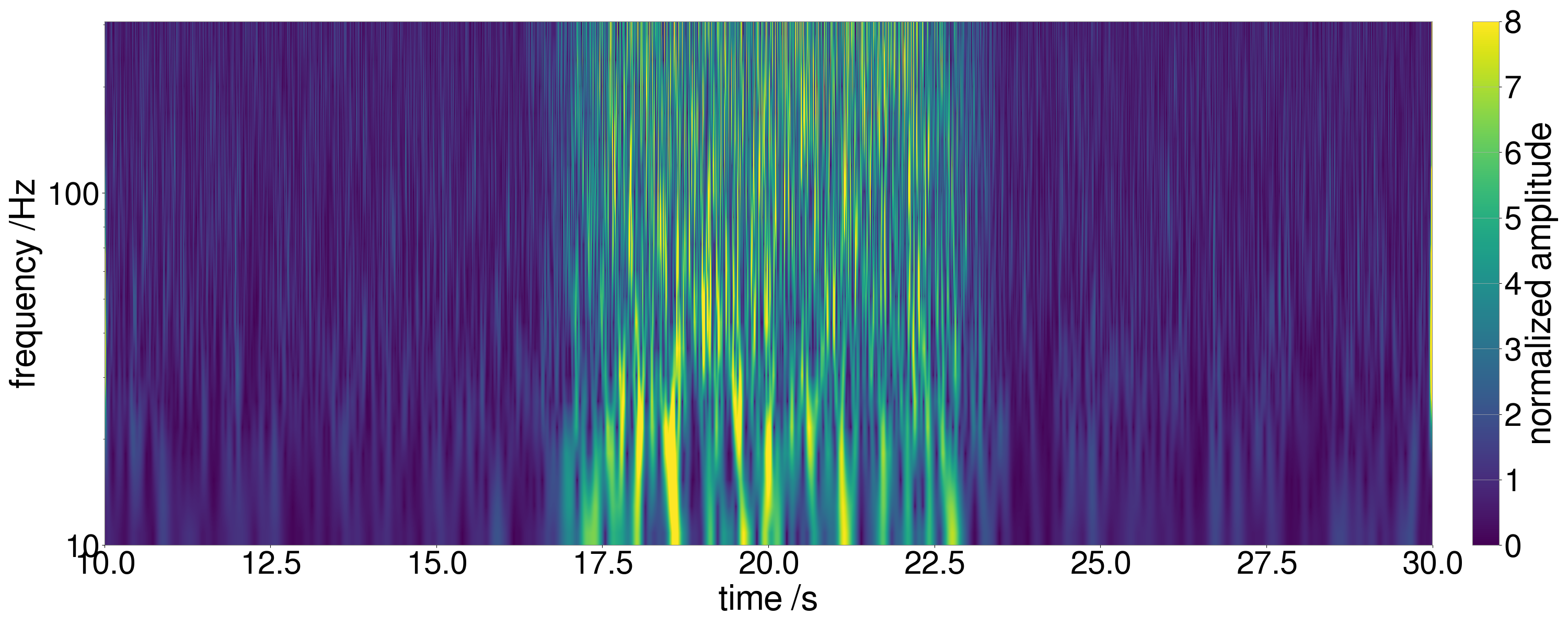}
        \caption{Time-frequency spectrogram of Gaussian noise with an 8-second Tukey window applied at 16 to 24 seconds.}
        \label{fig:QScan_GaussianTukey}
    \end{subfigure}
    \begin{subfigure}[b]{0.475\textwidth}
        \includegraphics[width=\textwidth]{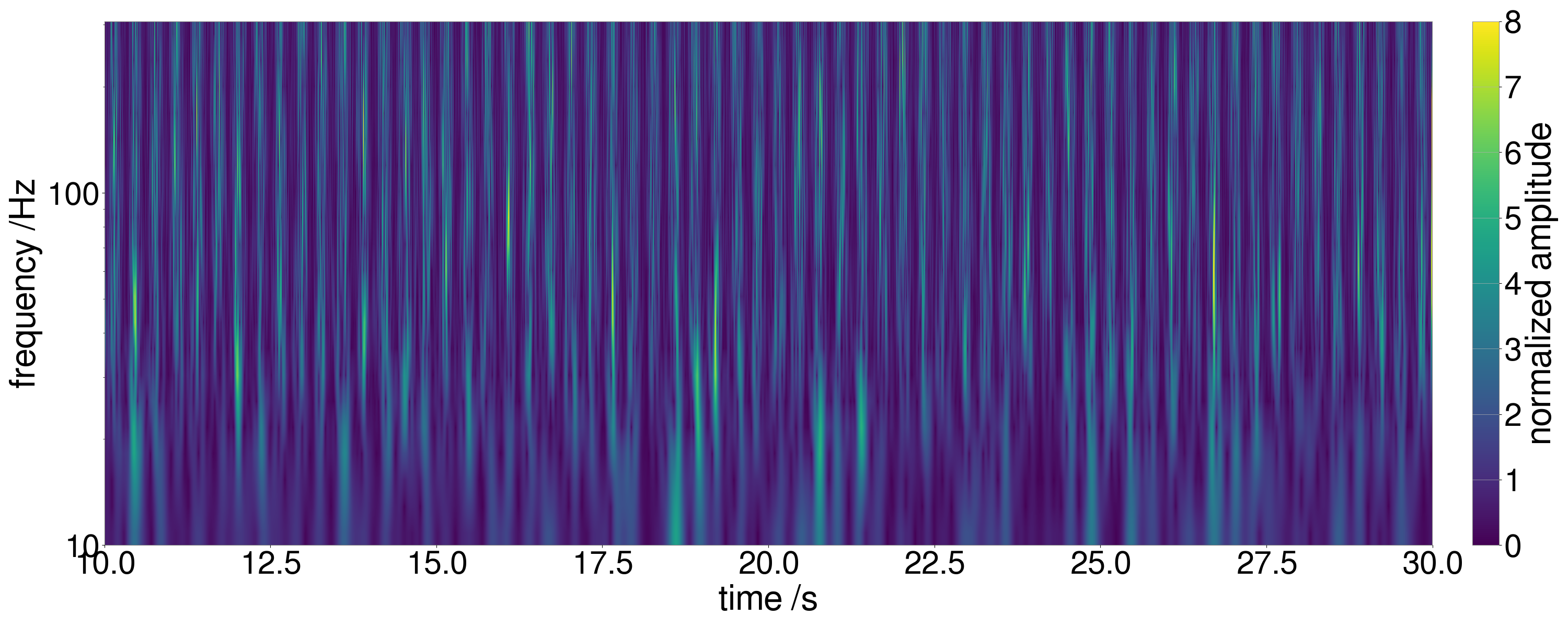}
        \caption{Time-frequency spectrogram of Gaussian noise with an oscillating amplitude of 3.2 Hz.}
        \label{fig:QScan_GaussianModulation}
    \end{subfigure}
    \caption{Time-frequency representation for three artificially generated timeseries using the Q-transform \cite{Chatterji_2004}. Note that all plots have been restricted to the period between 10 and 30 seconds to better highlight the oscillating peaks in plot \subref{fig:QScan_GaussianModulation}.}
    \label{fig:QScan_Comparison}
\end{figure}

The data and time-frequency spectrograms for one sample of each type of noise are plotted in Figure \ref{fig:Strain_Comparison} and Figure \ref{fig:QScan_Comparison} respectively.

Model A is almost identical to the stationary noise in both of these plots, except for the spike in power created by the Tukey window. This is exactly as we would expect given the noise is the sum of two separate Gaussian timeseries for this period ($n_1(t)$ and $n_2(t)$ in Equation \eqref{eq:non-stationary_form}, where the contribution of $n_2(t)$ is 0 for the rest of the time due to the amplitude $B(t)$).

The power spectra for each of these models was taken using the Welch method. Consequently, this means that the spectra we have calculated correspond to the stationary part of the spectrum. The standard procedure in searches is to estimate the power spectrum from longer stretches of data using the median \cite{Aasi_2013}. This difference should not change our conclusions, as they are comparable spectra for when the size of the non-stationarity is small, as is the case here.

\begin{figure}
    \centering
    \includegraphics[width=0.45\textwidth]{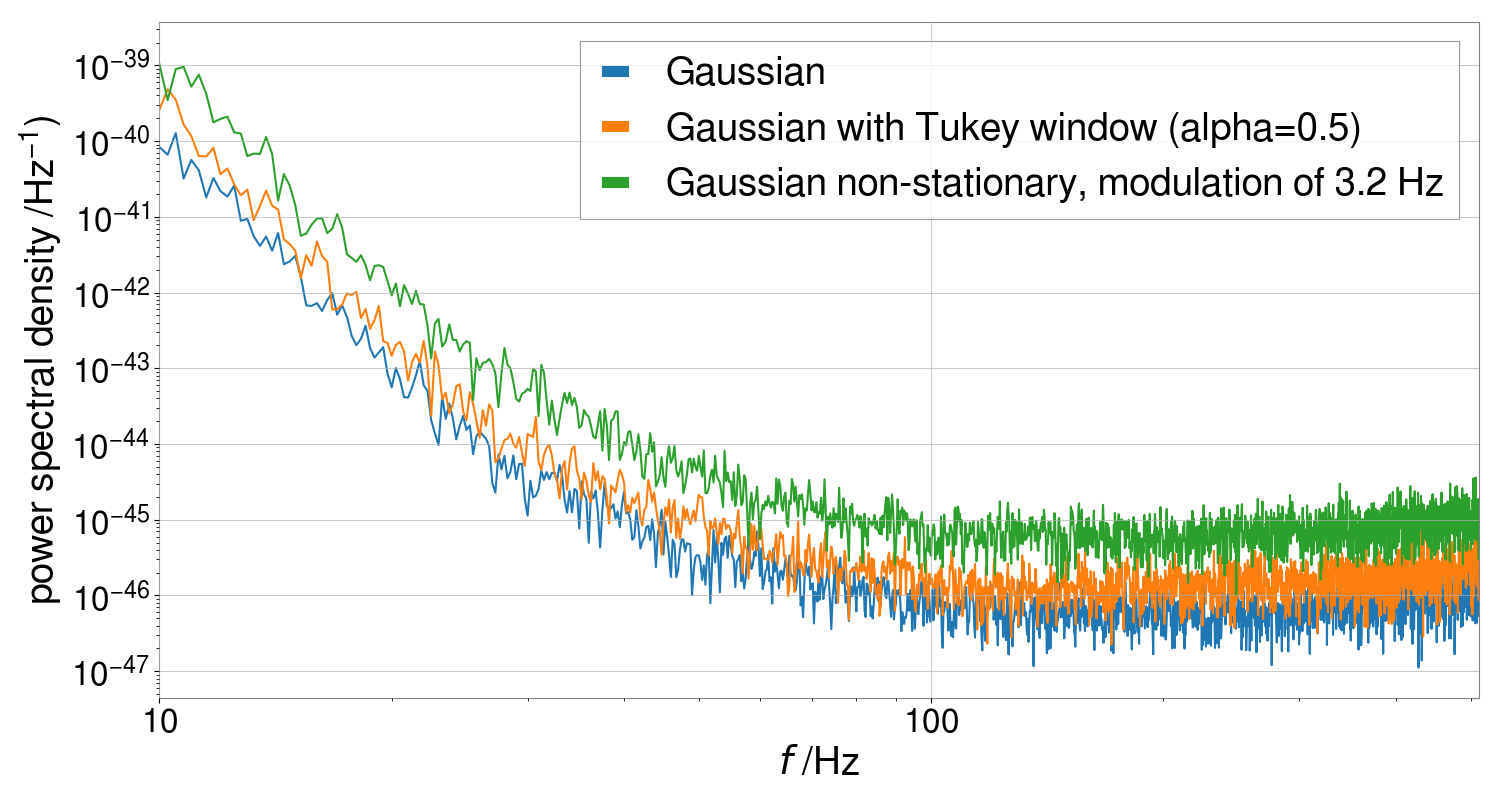}
    \caption{Power spectrum comparing 40-second samples of stationary Gaussian noise (blue), Gaussian noise made non-stationary by a Tukey window (orange), and Gaussian noise made non-stationary by a sinusoidal modulation of 3.2 Hz (green).}
    \label{fig:PSD_Comparison}
\end{figure}

When comparing the power spectra to each other (see Figure \ref{fig:PSD_Comparison}), the curve for Model A is very similar to the stationary Gaussian case. This is expected, as the power of the 8 seconds of non-stationarity would be washed out by the remaining 32 seconds of stationary Gaussian data.

While the spectrum for Model A has on average 2.8 times the power of the stationary Gaussian model, the spectrum for Model B has on average 13.3 times more power, an entire order of magnitude higher than the stationary Gaussian model. This is because the amplitude $B(t)$ extends over the entire 40-second sample, rather than a fraction of the entire duration. This means that the excess power will not be washed out by the stationary part.

So far, this has been a very qualitative review of the differences between these models. To get a better understanding of the deviations from stationarity, we then calculated the noise covariance for each of these models using the formalism described by Section \ref{sec:effect_of_nonst}.

\begin{figure}
    \begin{subfigure}[b]{0.4\textwidth}
        \centering
        \includegraphics[width=\textwidth]{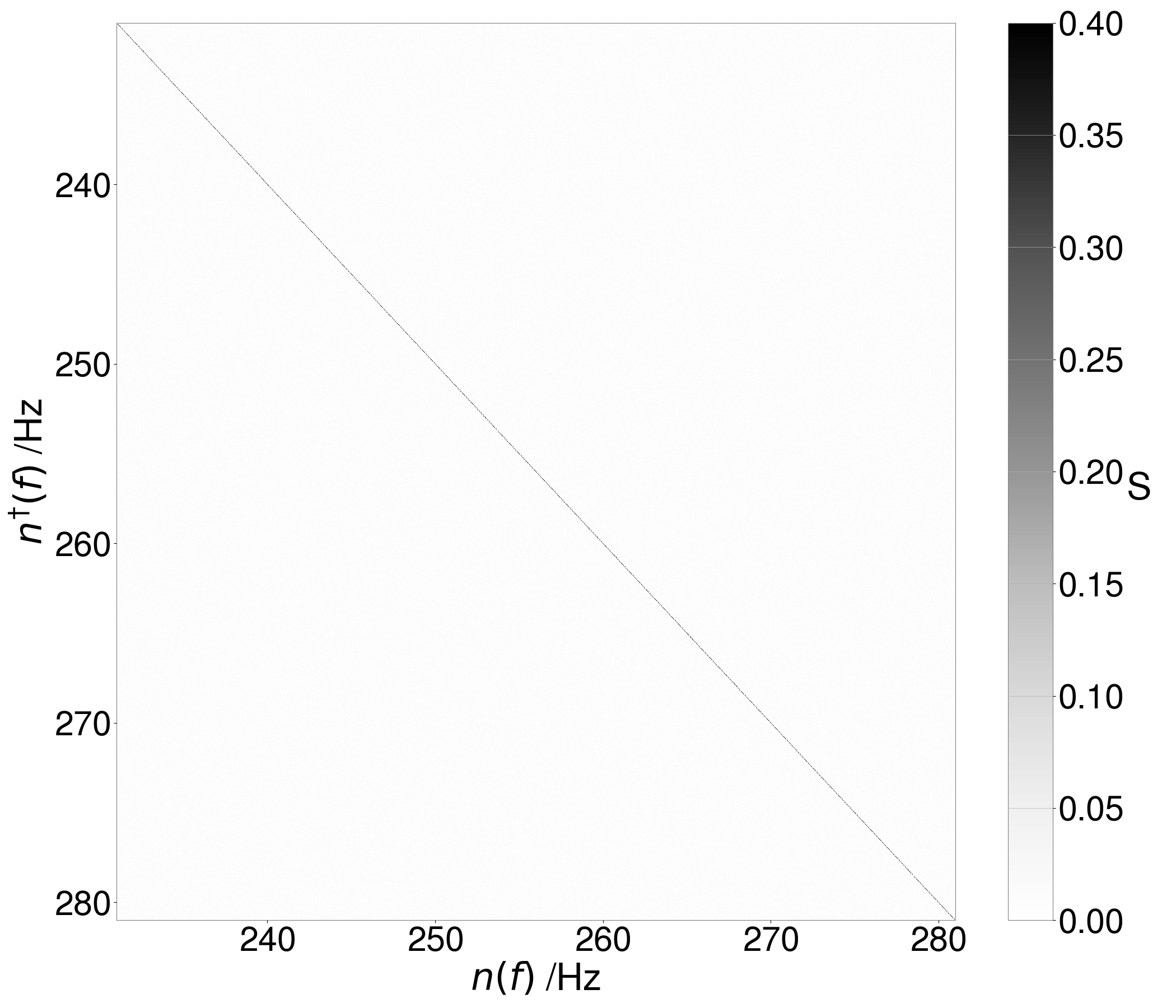}
        \caption{Noise covariance for Gaussian data.}
        \label{fig:Noise_Covariance_Gaussian}
    \end{subfigure}
    \begin{subfigure}[b]{0.4\textwidth}
        \centering
        \includegraphics[width=\textwidth]{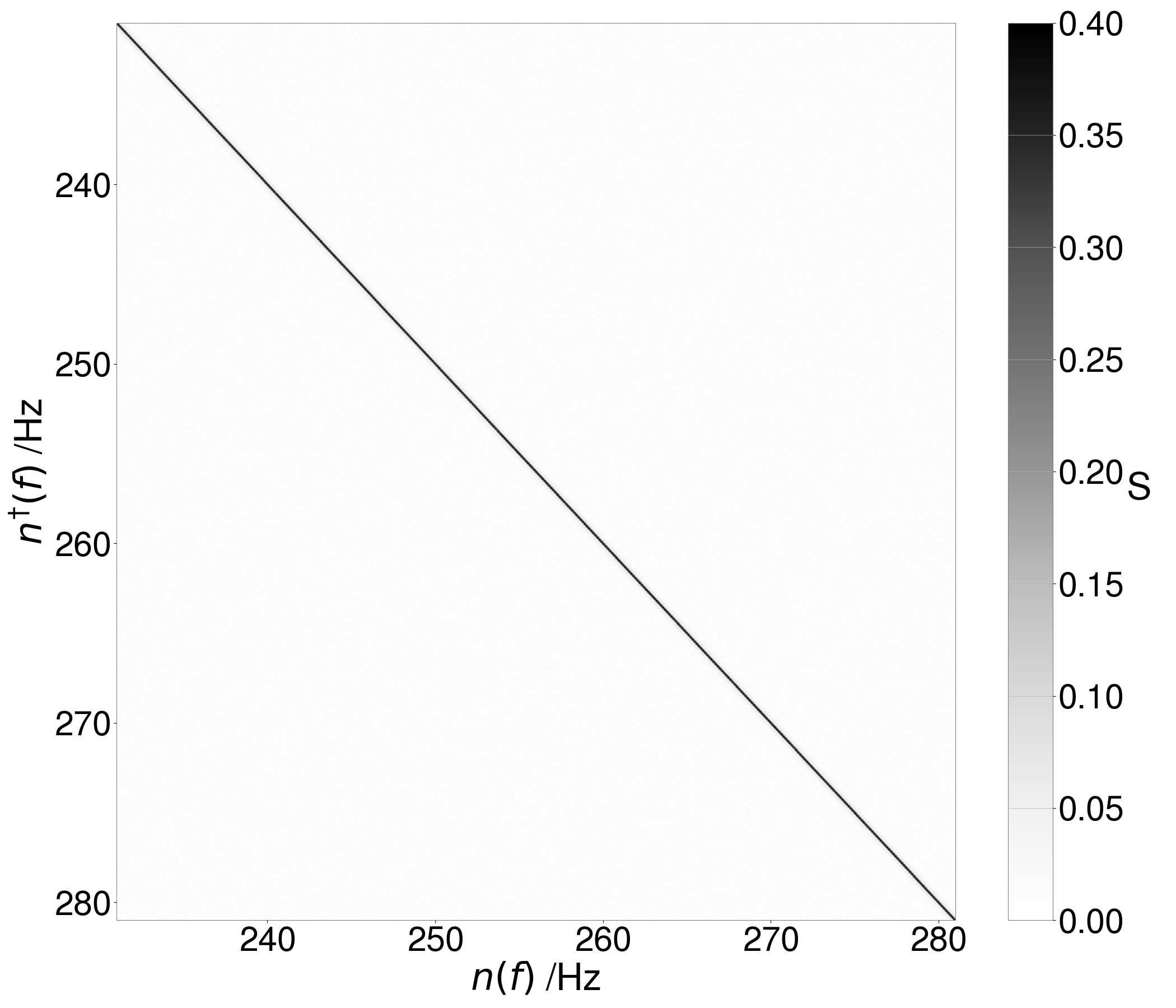}
        \caption{Noise covariance for Gaussian data with an 8-second Tukey window applied at 16 to 24 seconds. Features to be noticed are several sets of off-diagonal lines clustered very close to the leading diagonal.}
        \label{fig:Noise_Covariance_GaussianTukey}
    \end{subfigure}
    \begin{subfigure}[b]{0.4\textwidth}
        \centering
        \includegraphics[width=\textwidth]{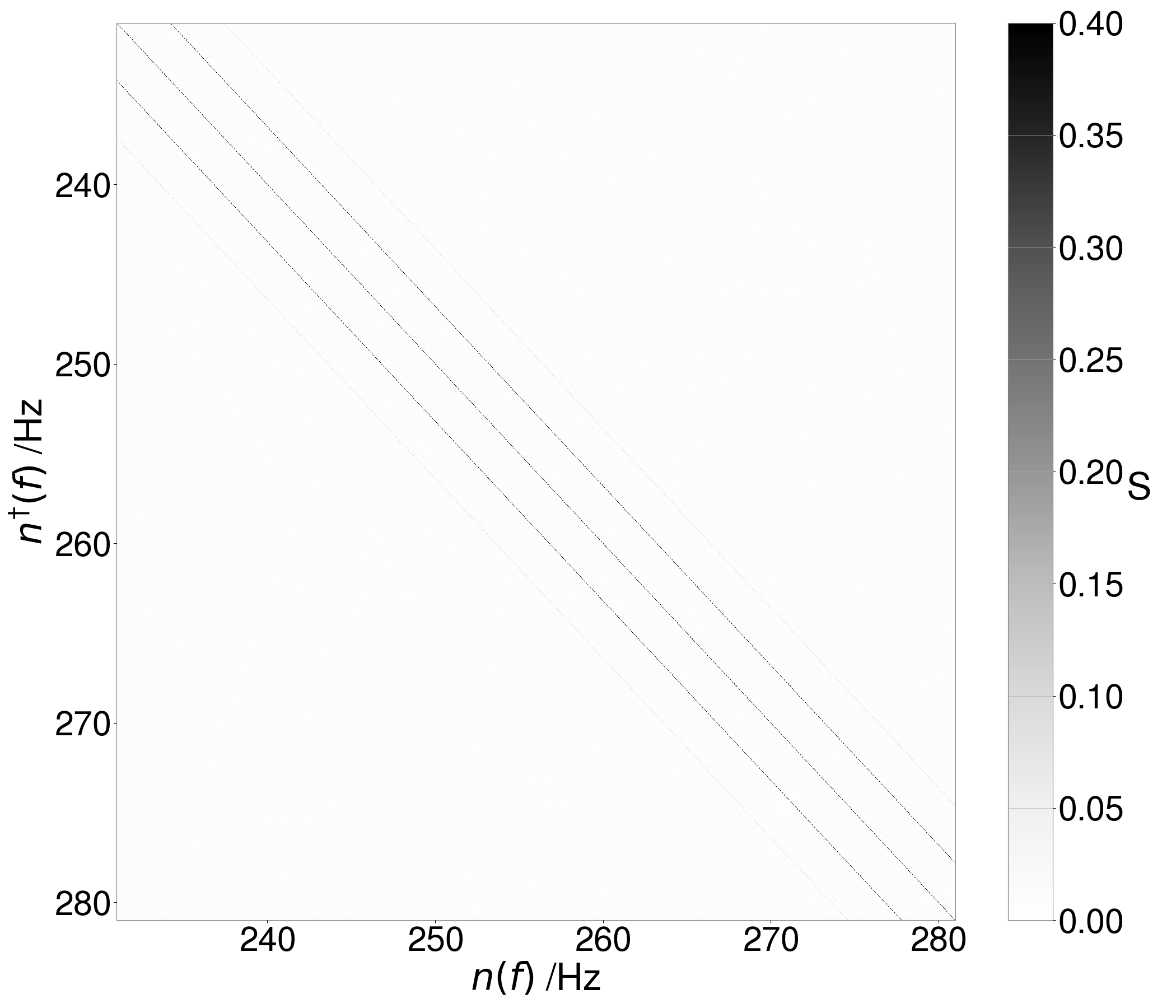}
        \caption{Noise covariance for Gaussian data with an oscillating amplitude of 3.2 Hz. Features to be noticed are the two sets of off-diagonal lines, where the second set is only just visible.}
        \label{fig:Noise_Covariance_GaussianModulation}
    \end{subfigure}
    \caption{Noise covariance for three artificially generated timeseries. Note that data has been whitened and rescaled so that all values are between 0 and 1, for better comparison. We further restrict the colour axis to only values between 0 and 0.4, to better highlight the off-diagonal terms in plot \subref{fig:Noise_Covariance_GaussianModulation}.}
    \label{fig:Noise_Covariance_Comparison}
\end{figure}

According to Equation \eqref{eq:noise_covariance_non-stationary}, the noise covariance is calculated as the inner product of the noise $n$ with its Hermitian conjugate $n^\dagger$ averaged over infinite noise realisations. Here we are realistically limited to 10,000 realisations. The resulting covariance matrices are plotted in Figure \ref{fig:Noise_Covariance_Comparison}, limited to the 50 Hz around 256 Hz for both the $x$- and $y$-axis to better showcase the diagonal and non-diagonal terms.

The stationary Gaussian model is a diagonal line of constant value, with every off-diagonal term on average 100 times smaller than the diagonal terms. This is as we predicted in Equation \eqref{eq:noise_covariance_stationary}, and we would expect the off-diagonal terms to be exactly zero if we were able to average over infinite noise realisations, rather than the 10,000 we were realistically limited to. The difference between Model A and the stationary Gaussian noise model is evident by the thicker width of the leading diagonal.

Model B also has a diagonal line of constant value, in addition to off-diagonal lines. It is these lines that represent the deviation from stationarity, with their distance from the central diagonal line related to the amplitude $B(t)$. The rest of the values are approximately zero.

While it is not immediately apparent in the plots shown, there is not just one set of off-diagonal lines created by the non-stationarity. The values of the first pair of lines off the central diagonal line are about a third of those on the central diagonal. The second set is just visible in Figure \ref{fig:Noise_Covariance_GaussianModulation}, with the values close to a tenth of those on the leading diagonal. Values in the other sets of lines are so small as to not be discernible from the zero values. The magnitude of the values and the distance of the lines in relation to the leading diagonal is related to the strength of the non-stationarity. In Figure \ref{fig:Noise_Covariance_GaussianModulation}, that refers to the frequency at which $B(t)$ oscillates.

In Section \ref{ssec:noise_characterisation}, we stated that Gaussian noise can be completely characterised with its noise covariance, and it should now be apparent that the off-diagonal terms are how the non-stationarity manifests in the noise covariance matrix. It is these non-diagonal terms that prevent us from assuming a simpler likelihood form. However, by knowing the noise covariance, we can calculate the extent that we will mismodel the data. As such, just by knowing $\Sigma$, we have a great description of the data. Unfortunately, estimating $\Sigma$ accurately is very difficult; we have tried to minimise variance in the data by taking the average over 10,000 realisations of the noise, but this is obviously not possible for real noise.

We now look at the discrepancies in the covariance matrix of gravitational-wave waveform parameters that we might expect to see for mergers when we incorrectly assume a stationary noise covariance matrix.

\subsection{The Effect of Non-Stationarity on the Covariance Matrix of a Short Gravitational-Wave Signal} \label{ssec:example_of_effect_of_nonst_for_BBH}

We model a waveform with parameters purposefully chosen to be similar to GW150914 \cite{Abbott_2016}, with SNR set to 15. This is because the waveform is short in duration, and representative of the signals that LIGO can currently detect. That means we can show how uncertain the parameters of signals we currently observe will be given the non-stationarity is not accounted for.

The covariance matrix for the waveform parameters shows how confidently we can measure each parameter. We will compare the covariance matrix of the waveform parameters for Model A and Model B with the stationary Gaussian model. This was chosen as the reference because stationary Gaussian noise perfectly matches the likelihood form used, and so the posteriors will take on their true form.

The waveform model we have chosen to use is TaylorF2, as defined by Reference \cite{Poisson_1995}, in which the signal $h(f)$ is of the form
\begin{equation}
    h(f) = f^{-\frac{7}{6}} e^{\mathcal{A} + i \psi\left( f \right)} \label{eq:waveform},
\end{equation}
where to 1.5PN,
\begin{dmath}
    \psi(f) = 2 \pi f t_c + \phi_c - \frac{\pi}{4} + \frac{3}{128} \left(  \pi \mathcal{M} f \right)^{-\frac{5}{3}} \left(  1 + \frac{20}{9} \left(  \frac{743}{336} + \frac{11}{4} \eta \right) \left( \pi M f \right)^{\frac{2}{3}} - 4 \left( 4 \pi - \beta \right) \left( \pi M f \right) + 10 \varepsilon \left(  \frac{3058673}{1016064} + \frac{5429}{1008} \eta + \frac{617}{144} \eta^2 - \sigma \right) \left( \pi M f \right)^{\frac{4}{3}} \right) \label{eq:psi}.
\end{dmath}

While we are using parameters for a binary black hole here, we believe that TaylorF2 can be used to sufficiently show the impact that non-stationarity can have on parameter estimation. We are not comparing different waveform models, but instead are comparing the effects of non-stationary noise models with a stationary noise model, using the same waveform throughout.

For simplicity, we reduce the TaylorF2 waveform to a form that only uses the parameters $\mathcal{A}$, $\phi$, $t_c$, and $\mathcal{M}$. With the exception of the chirp mass $\mathcal{M}$, each of these parameters is linearised, and so their corresponding values in the covariance matrix can be interpreted easily. The mass component can also be linearised by defining
\begin{equation}
    \lambda_m = \frac{3}{128} \left( \pi \mathcal{M} f_0 \right)^{-\frac{5}{3}} \label{chirp_mass}.
\end{equation}
This is the chirp time, the time it takes the waveform to go from $f_0$ to coalescence. Hence, we can re-express $\psi(f)$ as
\begin{equation}
    \psi(f) = 2 \pi f t_c + \phi_c - \frac{\pi}{4} + \lambda_m \left(  \frac{f}{f_0} \right)^{-\frac{5}{3}} \left( 1 + \dots \right).
\end{equation}

Using this waveform and the noise covariance matrices just found (and plotted in Figure \ref{fig:Noise_Covariance_Comparison}, the covariance matrix of the parameters can be calculated using Equation \eqref{eq:covariance_matrix_non-stationary}. The fractional difference between the covariance matrices of these two models and the stationary Gaussian covariance matrix are shown in Tables \ref{tab:Covariance_GaussianTukey} and \ref{tab:Covariance_GaussianModulation}. Note that the cross-terms of $\Delta \mathcal{A}$ and $\Delta \phi$ will always be 0 as the parameters are complex counterparts. We also reiterate that it does not make sense to compare the results of Model A directly with the results of Model B because they are very different models of noise.

\begin{table}[b]
    \caption{Fractional difference between the covariance matrix of Gaussian noise with an 8-second Tukey window, and stationary Gaussian noise.}
    \label{tab:Covariance_GaussianTukey}
    \begin{ruledtabular}
        \begin{tabular}{r | *{4}{c}}
            \enskip & $\Delta \mathcal{A}$ & $\Delta \phi$& $\Delta \lambda_m$ & $\Delta t_c$ \\
        \hline
            $\Delta \mathcal{A}$ & 4.008 & 0.0 & 4.008 & 4.007 \\
            $\Delta \phi$ & 0.0 & 4.008 & -0.5674 & -4.401 \\
            $\Delta \lambda_m$ & 4.008 & -0.5674 & 4.007 & 4.007 \\
            $\Delta t_c$ & 4.007 & -4.401 & 4.007 & 4.001
        \end{tabular}
    \end{ruledtabular}
\end{table}

\begin{table}[b]
    \caption{Fractional difference between the covariance matrix of Gaussian noise with an oscillating amplitude of 3.2 Hz, and stationary Gaussian noise.} \label{tab:Covariance_GaussianModulation}
    \begin{ruledtabular}
        \begin{tabular}{r | *{4}{c}}
            \enskip & $\Delta \mathcal{A}$ & $\Delta \phi$& $\Delta \lambda_m$ & $\Delta t_c$ \\
        \hline
            $\Delta \mathcal{A}$ & 1.32 & 0.0 & 1.299 & 1.35 \\
            $\Delta \phi$ & 0.0 & 1.32 & 1.147 & 1.018 \\
            $\Delta \lambda_m$ & 1.299 & 1.147 & 1.27 & 1.342 \\
            $\Delta t_c$ & 1.35 & 1.018 & 1.342 & 1.332
        \end{tabular}
    \end{ruledtabular}
\end{table}

The fractional differences in each value of the covariance matrix in Table \ref{tab:Covariance_GaussianModulation} are fairly consistent, and relatively small. As the non-stationarity extends over the entire sample, that means the extra power is also spread over the sample, and so does not greatly compromise parameter estimation results for a short signal.

Conversely, Model A was chosen to have the non-stationarity focused into a single section of the data, centered around the coalescence of the waveform. Bearing in mind the short duration of the waveform being considered, this effectively means we have twice as much stationary noise around the waveform than for the stationary Gaussian model. This is reflected by the fact that the values of the covariance matrix in Table \ref{tab:Covariance_GaussianTukey} are in general four times as large as for the stationary Gaussian model. This would mean that modelling this data as stationary Gaussian would produce much less confident posteriors than if the correct likelihood form was used.

One final point to note with Tables \ref{tab:Covariance_GaussianTukey} and \ref{tab:Covariance_GaussianModulation} is the presence of several values of the covariance for which the fractional difference is negative. This does not mean that the non-stationary noise produces a more accurate parameter in these cases. Recall that the covariance matrix predicts how well we can measure a parameter, and that the posterior is only correctly found when the likelihood form matches the data. Equation \eqref{eq:theta_av} tells us that the posterior will still be centered around the same point, so any mismodelling will only come from the width of the posterior. That means that the best estimate we will get of the parameters will come from stationary Gaussian data, and the non-stationary models will mean the uncertainties of the parameters could be overestimated or underestimated. The negative values correspond to an underestimate of the uncertainties.

We have shown that modelling non-stationary data as stationary will always affect the confidence with which parameters are estimated. However, in cases where the non-stationary period is of a comparable length to the noise sample duration, but is relatively small in amplitude, the extent to which short BBH-like signals are under- or overestimated is minimal. The same cannot be said for a loud spike of non-stationarity as in Model A. The limit at which a model can no longer be approximated as stationary is a topic of future research interest. We also reiterate that these results were generated as the average of 10,000 realisations of noise; when considering real interferometer data, we would only be able to use one sample, and so would expect the non-stationary results to be even worse, as the off-diagonal terms of the noise covariance will not average out.

\subsection{The Effect of Non-Stationarity on the Covariance Matrix of a Long Gravitational-Wave Signal} \label{ssec:example_of_effect_of_nonst_for_BNS}

The majority of signals detectors currently observe are only a few seconds long. This means that extended periods of non-stationarity should not cause a significant problem. However, it will become increasingly important to handle the effect of non-stationarity when detectors such as the Einstein Telescope begin observing. These are expected to detect much longer signals \cite{Regimbau_2012}, which put stringent requirements on stationarity. The analysis problems inherent to the Einstein Telescope are well documented (for example, \cite{Regimbau_2012, Bosi_2010, Punturo_2010}).

We look now at how a longer signal might be affected by the non-stationarity by comparing the covariance matrices seen in Tables \ref{tab:Covariance_GaussianTukey} and \ref{tab:Covariance_GaussianModulation}, but now considering a signal modelled after GW190814 \cite{Abbott_2020_GW190814}. This model was chosen due to the large mass difference and relatively small masses resulting in a merger of duration of tens of seconds. The effect of the mismodelling would be even more significant in an even longer signal, like we would see from a BNS-like object. However, we cannot inject a BNS-like signal into 40 seconds of data, as the signal is of a similar duration, and we would experience wraparound effects in the frequency-domain that would in turn distort the results. On the other hand, analysing data longer than 40 seconds would be both a challenge computationally, and make the results more difficult to compare to the shorter signal. 

\begin{table}[b]
    \caption{Fractional difference between the covariance matrix of Gaussian noise with an 8-second Tukey window, and stationary Gaussian noise for a GW190814-like object.}
    \label{tab:Covariance_GaussianTukey_BNS}
    \begin{ruledtabular}
        \begin{tabular}{r | *{4}{c}}
            \enskip & $\Delta \mathcal{A}$ & $\Delta \phi$& $\Delta \lambda_m$ & $\Delta t_c$ \\
        \hline
            $\Delta \mathcal{A}$ & 3.362 & 0.0 & 0.7657 & 4.073 \\
            $\Delta \phi$ & 0.0 & 3.362 & -5.171 & -21.54 \\
            $\Delta \lambda_m$ & 0.7657 & -5.171 & 1.956 & 3.907 \\
            $\Delta t_c$ & 4.073 & -21.54 & 3.907 & 4.001
        \end{tabular}
    \end{ruledtabular}
\end{table}

\begin{table}[b]
    \caption{Fractional difference between the covariance matrix of Gaussian noise with an oscillating amplitude of 3.2 Hz, and stationary Gaussian noise for a GW190814-like object.} \label{tab:Covariance_GaussianModulation_BNS}
    \begin{ruledtabular}
        \begin{tabular}{r | *{4}{c}}
            \enskip & $\Delta \mathcal{A}$ & $\Delta \phi$& $\Delta \lambda_m$ & $\Delta t_c$ \\
        \hline
            $\Delta \mathcal{A}$ & 1.234 & 0.0 & 1.965 & 0.8001 \\
            $\Delta \phi$ &  0.0 & 1.234 & -1.838 & 1.885 \\
            $\Delta \lambda_m$ & 1.965 & -1.838 & 1.219 & -0.1868 \\
            $\Delta t_c$ & 0.8001 & 1.885 & -0.1868 & 0.3767
        \end{tabular}
    \end{ruledtabular}
\end{table}

Just as we saw in Section \ref{ssec:example_of_effect_of_nonst_for_BBH}, both the short period of non-stationarity in Model A and the extended period of non-stationarity in Model B have a detrimental effect on the estimated parameters of the waveform. An interesting comparison would be between Tables \ref{tab:Covariance_GaussianTukey} and \ref{tab:Covariance_GaussianTukey_BNS}, and \ref{tab:Covariance_GaussianModulation} and \ref{tab:Covariance_GaussianModulation_BNS}. This would allow us to compare how much more greatly affected a long signal would be affected by non-stationarity than a shorter signal.

For Model B, the fractional difference in covariances for GW190814 are approximately the same as for the GW150914-like signal, with the greatest difference being the increase in the magnitude of the cross-terms (such as for $\Delta \mathcal{A} \Delta \lambda_m$). For Model A, the majority of the covariances are slightly smaller, but still of the same order of magnitude. The greatest discrepancy comes from the underestimate of the $\Delta \phi \Delta t_c$ term. This arises from the length increase of the signal, which now coalesces in the centre of the Tukey window, but the inspiral starts before the window, transitioning between two radically different forms of the noise. We expect to see even greater discrepancies with the stationary model when we consistently observe longer signals, a problem that has already been identified for LISA \cite{Edwards_2020}, for which antenna repointing will create data gaps during the observation of signals. The assumption of stationarity in these periods will lead to parameter estimation biases.

\subsection{Variation of the Fisher Matrix Over Time} \label{ssec:Fisher_over_tc}

Perhaps one of the most surprising things seen when comparing the results in Section \ref{ssec:example_of_effect_of_nonst_for_BBH} and \ref{ssec:example_of_effect_of_nonst_for_BNS} is the decrease in magnitude of the covariance matrix values as the waveform duration increased. This is because we calculate the quantity $\omega^\dagger S^{-1} \omega$, a Fisher matrix, and must invert it to find the covariance matrix. In the process of inverting the matrix, terms of the matrix are affected by the mass components, which are larger for larger values of $\mathcal{M}$. If we were to instead look at the Fisher matrix, we would not see the influence of these mass components on the other values. As a corollary, by comparing Fisher matrices, while we could evaluate the difference in measurability of the different parameters for the non-stationary noise models compared to the stationary noise model, it would not help us understand how greatly under- or overestimated the parameters might be, and an understanding of both is useful.

\begin{figure}
    \begin{subfigure}[b]{0.475\textwidth}
        \centering
        \includegraphics[width=\textwidth]{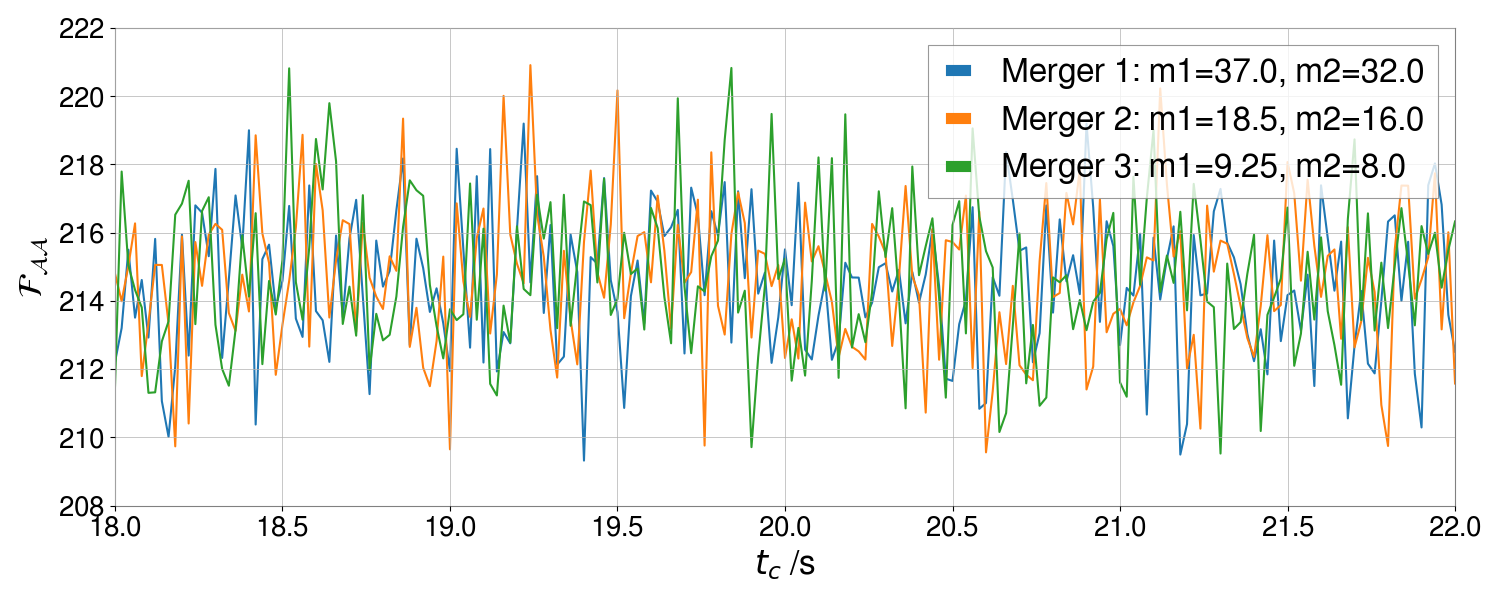}
        \caption{Values of $\mathcal{F}_{\mathcal{A}\mathcal{A}}$ at times $t_c$ for three waveforms injected into Gaussian data.}
        \label{fig:Fisher_Matrix_Over_tc_Gaussian}
    \end{subfigure}
    \begin{subfigure}[b]{0.475\textwidth}
        \centering
        \includegraphics[width=\textwidth]{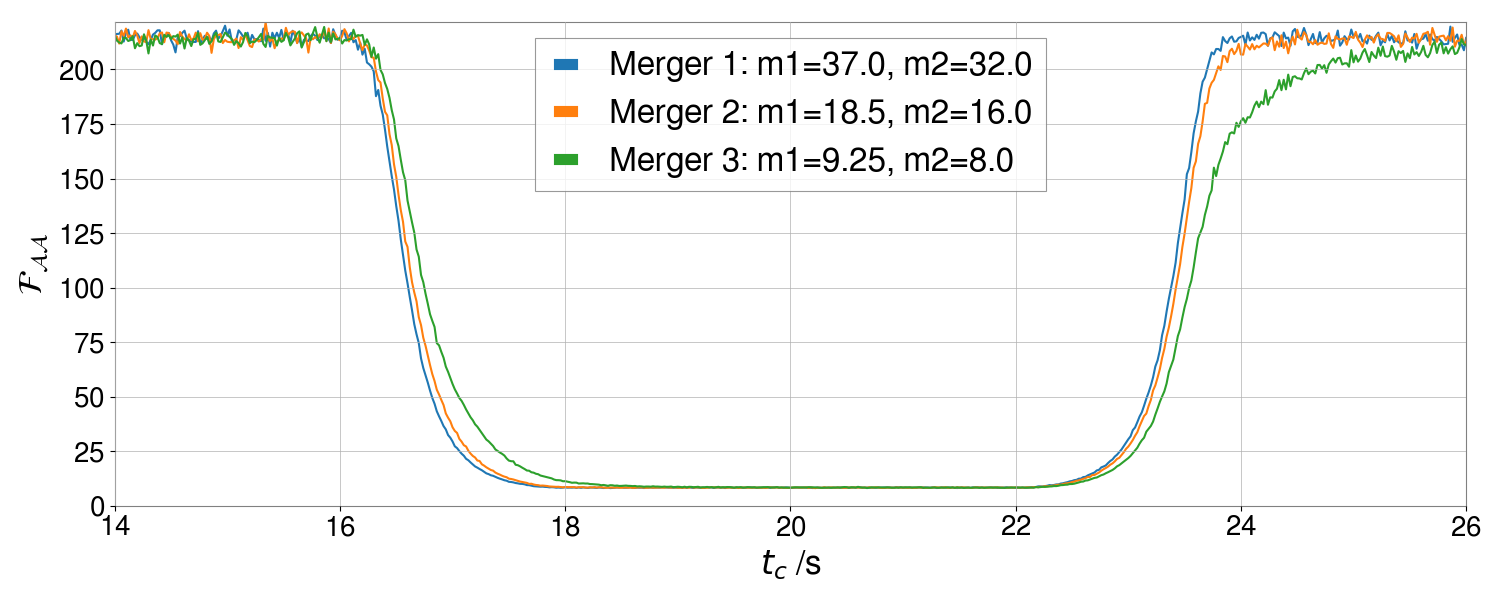}
        \caption{Values of $\mathcal{F}_{\mathcal{A}\mathcal{A}}$ at times $t_c$ for three waveforms injected into Gaussian data with an 8-second Tukey window applied at 16 to 24 seconds.}
        \label{fig:Fisher_Matrix_Over_tc_GaussianTukey}
    \end{subfigure}
    \begin{subfigure}[b]{0.475\textwidth}
        \centering
        \includegraphics[width=\textwidth]{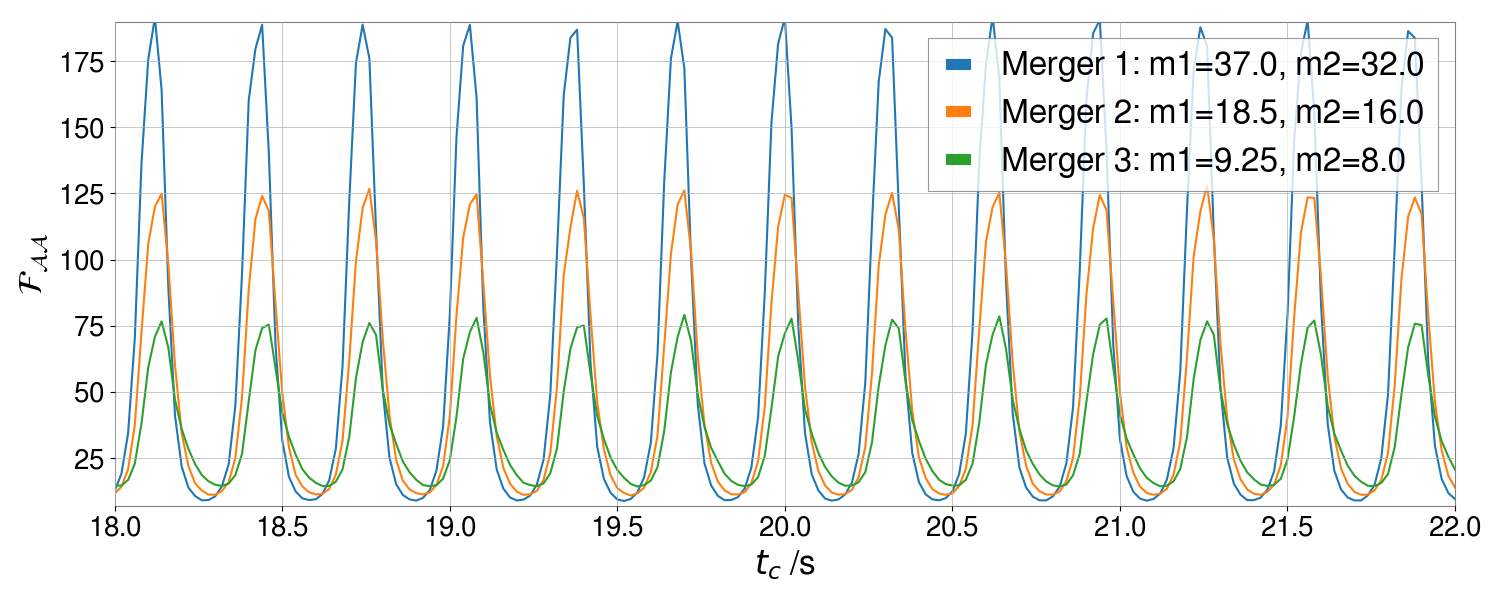}
        \caption{Values of $\mathcal{F}_{\mathcal{A}\mathcal{A}}$ at times $t_c$ for three waveforms injected into Gaussian data with an oscillating amplitude of 3.2 Hz.}
        \label{fig:Fisher_Matrix_Over_tc_GaussianModulation}
    \end{subfigure}
    \caption{Values of $\mathcal{F}_{\mathcal{A}\mathcal{A}}$ for different values of coalescence time $t_c$ for three waveforms where only their total masses are varied. Note that masses are in terms of solar mass. The $t_c$ range for panel \subref{fig:Fisher_Matrix_Over_tc_GaussianModulation} is displayed from 4 to 16 seconds, to better showcase the effect of the entire Tukey window.}
    \label{fig:Fisher_Matrix_Over_tc}
\end{figure}

Because of the non-stationarity of the data, we do expect to see the exact values of the Fisher matrix vary with $t_c$. To explore to what extent the values vary with time, we have calculated the Fisher matrix at different values of $t_c$ for three different waveforms of decreasing total mass. The corresponding values of one parameter of these Fisher matrices, the amplitude-amplitude term $\mathcal{F}_{\mathcal{A}\mathcal{A}}$, is plotted in Figure \ref{fig:Fisher_Matrix_Over_tc}.

In the case where the noise is stationary Gaussian, the value of $\mathcal{F}_{\mathcal{A}\mathcal{A}}$ is roughly constant regardless of the waveform, as we would expect, with the small fluctuations we do see arising from the fact we could only average over 10,000 realisations of noise, rather than an infinite number of realisations.

At times before and after the Tukey window is applied to Model A ($16<t_c<24$s), the values of $\mathcal{F}_{\mathcal{A}\mathcal{A}}$ are comparable to the stationary Gaussian noise model, because it simply is stationary Gaussian noise at these times. In the period when the Tukey window is applied, the values of $\mathcal{F}_{\mathcal{A}\mathcal{A}}$ are still roughly constant for all waveforms, although the values are much smaller. This is consistent with the assertion that the increased level of noise in the Tukey window means that the parameters are less measurable. We also note that the line corresponding to Merger 3 (between masses $m_1=9.25 M_{\odot}$ and $m_2 = 8.0 M_{\odot}$) has a greater offset than for the other two mergers. The longer waveform of Merger 3 averages over more time, and so the Fisher matrix values shift in time.

The sinusoidal modulation is visible in the results for Model B. We can quite clearly see here that the parameters become more measurable (the values for $\mathcal{F}_{\mathcal{A}\mathcal{A}}$ spikes) whenever the oscillating amplitude is at a minimum. Again, we can see a difference caused by the duration of the signals; in this case, the very short duration of Merger 1 means that the signal is either entirely in or outside the spikes of non-stationarity, but the longer signals will extend outside of the spikes, and so the values of the Fisher matrix will average out in time. While not as tightly spaced as for the stationary Gaussian model, all the values are in the same order of magnitude, and so are still comparable to each other.

The non-stationarity described here is far more extreme than we would expect to see in real experiments, chosen instead to demonstrate the effect on the waveform parameters. Even so, the comparable measurability of the waveforms shows that the estimation of the parameters depends very strongly on when the merger takes place relative to the noise, more than the properties of the waveform itself.

\section{Conclusion and Future Work} \label{sec:conclusion}

We have undertaken one of the first analytical studies of the effect of mismodelling gravitational-wave strain data as stationary in parameter estimation. We established that mismodelling the data in this way will not bias the parameters, but will affect the width of the posterior. This means that there will be an under- or overestimate of the parameter credible intervals.

We demonstrated a methodology for investigating the effect of non-stationarity through the calculation of the noise covariance $\Sigma$, and the covariance of the waveform parameters. Using this methodology, we showed that any form of non-stationarity will affect the results of parameter estimation, but the credible intervals are more greatly misrepresented when the non-stationarity increases the amplitude of the noise, rather than the form of the non-stationarity itself; Reference \cite{Berry_2015} described similar results for a given SNR. The effect non-stationarity has on parameter estimation is more pronounced in the case when the signal itself is longer and extends over several forms of the noise, as parameters become harder to measure, as signified by the decrease in magnitude of the Fisher matrix values. This indicates that non-stationarity is a particular obstacle in reliably estimating the parameters of a BNS. Additionally, if we continue to assume the noise is Gaussian, all we need to know about the noise is the matrix $\Sigma$, since this square matrix completely characterises the data, with any non-stationarity manifesting in off-diagonal terms.

Now that a method for determining the covariance of stationary and non-stationary Gaussian noise has been determined, we believe there is merit in creating a measure of the effect of non-stationarity by comparing covariance matrices. We proposed that such a metric would concern deviations from the expected stationary Gaussian covariance we have obtained from the method above, with the main aim of establishing the point at which non-stationarity becomes a problem.

We believe it would be possible to predict the extent to which the posterior will be over- or underestimated, and so a formalism to account for this without significantly increasing computational cost should be possible to include in parameter estimation codes. Until a method for handling the non-stationarity is developed, the incorrect credible intervals could also be a limiting factor of astronomical observations using gravitational waves, not least of which being an estimation of the Hubble parameter. For example, Reference \cite{Del_Pozzo_2014} proposed that gravitational-wave events can be used to obtain a measurement of $H_0$ with an uncertainty of only a few percent; the methodology outlined in our paper could be applied to ensure that the confidence intervals are not being underestimated. We would also be concerned that incorporating the uncertainty of the PSD measurement as suggested by Reference \cite{Biscoveanu_2020} would only serve to compromise the $H_0$ measurement. Our methodology could also be particularly important when detectors such as the Einstein Telescope begin observing, as these are expected to detect much longer signals \cite{Regimbau_2012}, which put stringent requirements on stationarity.

One of the major issues we foresee with this work is the calculation of $\Sigma$. Therefore, it would be useful to delve into methods for estimating $\langle n n^\dagger \rangle$ on average to accurately find $\Sigma$ in realistic data.

We also recall that real noise is not only non-stationary, it is also non-Gaussian. However, it is typically modeled as stationary Gaussian noise for the same computational efficiency reasons as mentioned in Section \ref{ssec:derivation}. Although the effect that the non-stationarity has on the posterior is being examined through the method in this paper and any follow-up work, the same method cannot be used for non-Gaussian data owing to the different form the likelihood takes. Instead, we see the need for future work to develop suitable techniques to measure the effect on parameter estimation of mismodelling non-Gaussian data.

From this work, it would be possible to determine how non-ideal a dataset can be without compromising parameter estimation, create a robust measure for the extent to which these datasets are affected, and investigate whether current inference codes can be modified to account for the effect without greatly increasing computational cost.

\begin{acknowledgments}
We thank Ian Harry and John Veitch for useful discussions and suggestions. The authors are grateful for computational resources provided by the LIGO Laboratory and supported by National
Science Foundation Grants PHY-0757058 and PHY-0823459. OE was supported by an STFC studentship. AL thanks the STFC for support
through the grant ST/S000550/1. LKN thanks the UKRI Future Leaders Fellowship for support through the grant MR/T01881X/1. This document has been assigned LIGO Laboratory document number LIGO-P2100001.
\end{acknowledgments}

\bibliography{References}

\begin{thebibliography}{72}%
\makeatletter
\providecommand \@ifxundefined [1]{%
 \@ifx{#1\undefined}
}%
\providecommand \@ifnum [1]{%
 \ifnum #1\expandafter \@firstoftwo
 \else \expandafter \@secondoftwo
 \fi
}%
\providecommand \@ifx [1]{%
 \ifx #1\expandafter \@firstoftwo
 \else \expandafter \@secondoftwo
 \fi
}%
\providecommand \natexlab [1]{#1}%
\providecommand \enquote  [1]{``#1''}%
\providecommand \bibnamefont  [1]{#1}%
\providecommand \bibfnamefont [1]{#1}%
\providecommand \citenamefont [1]{#1}%
\providecommand \href@noop [0]{\@secondoftwo}%
\providecommand \href [0]{\begingroup \@sanitize@url \@href}%
\providecommand \@href[1]{\@@startlink{#1}\@@href}%
\providecommand \@@href[1]{\endgroup#1\@@endlink}%
\providecommand \@sanitize@url [0]{\catcode `\\12\catcode `\$12\catcode
  `\&12\catcode `\#12\catcode `\^12\catcode `\_12\catcode `\%12\relax}%
\providecommand \@@startlink[1]{}%
\providecommand \@@endlink[0]{}%
\providecommand \url  [0]{\begingroup\@sanitize@url \@url }%
\providecommand \@url [1]{\endgroup\@href {#1}{\urlprefix }}%
\providecommand \urlprefix  [0]{URL }%
\providecommand \Eprint [0]{\href }%
\providecommand \doibase [0]{https://doi.org/}%
\providecommand \selectlanguage [0]{\@gobble}%
\providecommand \bibinfo  [0]{\@secondoftwo}%
\providecommand \bibfield  [0]{\@secondoftwo}%
\providecommand \translation [1]{[#1]}%
\providecommand \BibitemOpen [0]{}%
\providecommand \bibitemStop [0]{}%
\providecommand \bibitemNoStop [0]{.\EOS\space}%
\providecommand \EOS [0]{\spacefactor3000\relax}%
\providecommand \BibitemShut  [1]{\csname bibitem#1\endcsname}%
\let\auto@bib@innerbib\@empty
\bibitem [{\citenamefont {Aasi}\ \emph {et~al.}(2015)\citenamefont {Aasi} \emph
  {et~al.}}]{Aasi_2015}%
  \BibitemOpen
  \bibfield  {author} {\bibinfo {author} {\bibfnamefont {J.}~\bibnamefont
  {Aasi}} \emph {et~al.},\ }\bibfield  {title} {\bibinfo {title} {Advanced
  ligo},\ }\href {https://doi.org/10.1088/0264-9381/32/7/074001} {\bibfield
  {journal} {\bibinfo  {journal} {Classical and Quantum Gravity}\ }\textbf
  {\bibinfo {volume} {32}},\ \bibinfo {pages} {074001} (\bibinfo {year}
  {2015})},\ \Eprint {https://arxiv.org/abs/1411.4547} {1411.4547 [gr-qc]}
  \BibitemShut {NoStop}%
\bibitem [{\citenamefont {Abbott}\ \emph {et~al.}(2017)\citenamefont {Abbott}
  \emph {et~al.}}]{Abbott_2017}%
  \BibitemOpen
  \bibfield  {author} {\bibinfo {author} {\bibfnamefont {B.~P.}\ \bibnamefont
  {Abbott}} \emph {et~al.} (\bibinfo {collaboration} {The LIGO Scientific
  Collaboration and Virgo Collaboration}),\ }\bibfield  {title} {\bibinfo
  {title} {Gw170814: A three-detector observation of gravitational waves from a
  binary black hole coalescence},\ }\bibfield  {journal} {\bibinfo  {journal}
  {Physical Review Letters}\ }\textbf {\bibinfo {volume} {119}},\ \href
  {https://doi.org/10.1103/physrevlett.119.141101}
  {10.1103/physrevlett.119.141101} (\bibinfo {year} {2017}),\ \Eprint
  {https://arxiv.org/abs/1709.09660} {arXiv:1709.09660 [gr-qc]} \BibitemShut
  {NoStop}%
\bibitem [{\citenamefont {Akutsu}\ \emph {et~al.}(2020)\citenamefont {Akutsu}
  \emph {et~al.}}]{Akutsu_2020}%
  \BibitemOpen
  \bibfield  {author} {\bibinfo {author} {\bibfnamefont {T.}~\bibnamefont
  {Akutsu}} \emph {et~al.},\ }\href@noop {} {\bibinfo {title} {Overview of
  kagra: Calibration, detector characterization, physical environmental
  monitors, and the geophysics interferometer}} (\bibinfo {year} {2020}),\
  \Eprint {https://arxiv.org/abs/2009.09305} {arXiv:2009.09305 [gr-qc]}
  \BibitemShut {NoStop}%
\bibitem [{\citenamefont {Babak}\ \emph {et~al.}(2013)\citenamefont {Babak}
  \emph {et~al.}}]{Babak_2013}%
  \BibitemOpen
  \bibfield  {author} {\bibinfo {author} {\bibfnamefont {S.}~\bibnamefont
  {Babak}} \emph {et~al.},\ }\bibfield  {title} {\bibinfo {title} {Searching
  for gravitational waves from binary coalescence},\ }\href
  {https://doi.org/10.1103/physrevd.87.024033} {\bibfield  {journal} {\bibinfo
  {journal} {Physical Review D}\ }\textbf {\bibinfo {volume} {87}},\ \bibinfo
  {pages} {024033} (\bibinfo {year} {2013})},\ \Eprint
  {https://arxiv.org/abs/1208.3491} {arXiv:1208.3491 [gr-qc]} \BibitemShut
  {NoStop}%
\bibitem [{\citenamefont {Camp}\ and\ \citenamefont
  {Cornish}(2004)}]{Camp_2004}%
  \BibitemOpen
  \bibfield  {author} {\bibinfo {author} {\bibfnamefont {J.~B.}\ \bibnamefont
  {Camp}}\ and\ \bibinfo {author} {\bibfnamefont {N.~J.}\ \bibnamefont
  {Cornish}},\ }\bibfield  {title} {\bibinfo {title} {Gravitational wave
  astronomy},\ }\href {https://doi.org/10.1146/annurev.nucl.54.070103.181251}
  {\bibfield  {journal} {\bibinfo  {journal} {Annual Review of Nuclear and
  Particle Science}\ }\textbf {\bibinfo {volume} {54}},\ \bibinfo {pages} {525}
  (\bibinfo {year} {2004})}\BibitemShut {NoStop}%
\bibitem [{\citenamefont {Veitch}\ \emph {et~al.}(2015)\citenamefont {Veitch}
  \emph {et~al.}}]{Veitch_2015}%
  \BibitemOpen
  \bibfield  {author} {\bibinfo {author} {\bibfnamefont {J.}~\bibnamefont
  {Veitch}} \emph {et~al.},\ }\bibfield  {title} {\bibinfo {title} {Parameter
  estimation for compact binaries with ground-based gravitational-wave
  observations using the lalinference software library},\ }\bibfield  {journal}
  {\bibinfo  {journal} {Physical Review D}\ }\textbf {\bibinfo {volume} {91}},\
  \href {https://doi.org/10.1103/physrevd.91.042003}
  {10.1103/physrevd.91.042003} (\bibinfo {year} {2015}),\ \Eprint
  {https://arxiv.org/abs/1409.7215} {arXiv:1409.7215 [gr-qc]} \BibitemShut
  {NoStop}%
\bibitem [{\citenamefont {Abbott}\ \emph
  {et~al.}(2016{\natexlab{a}})\citenamefont {Abbott} \emph
  {et~al.}}]{Abbott_2016_GW150914}%
  \BibitemOpen
  \bibfield  {author} {\bibinfo {author} {\bibfnamefont {B.~P.}\ \bibnamefont
  {Abbott}} \emph {et~al.} (\bibinfo {collaboration} {The LIGO Scientific
  Collaboration and Virgo Collaboration}),\ }\bibfield  {title} {\bibinfo
  {title} {Properties of the binary black hole merger gw150914},\ }\bibfield
  {journal} {\bibinfo  {journal} {Physical Review Letters}\ }\textbf {\bibinfo
  {volume} {116}},\ \href {https://doi.org/10.1103/physrevlett.116.241102}
  {10.1103/physrevlett.116.241102} (\bibinfo {year} {2016}{\natexlab{a}}),\
  \Eprint {https://arxiv.org/abs/1602.03840} {arXiv:1602.03840 [gr-qc]}
  \BibitemShut {NoStop}%
\bibitem [{\citenamefont {van~der Sluys}\ \emph
  {et~al.}(2008{\natexlab{a}})\citenamefont {van~der Sluys} \emph
  {et~al.}}]{van_der_Sluys_2008}%
  \BibitemOpen
  \bibfield  {author} {\bibinfo {author} {\bibfnamefont {M.}~\bibnamefont
  {van~der Sluys}} \emph {et~al.},\ }\bibfield  {title} {\bibinfo {title}
  {Parameter estimation of spinning binary inspirals using markov chain monte
  carlo},\ }\href {https://doi.org/10.1088/0264-9381/25/18/184011} {\bibfield
  {journal} {\bibinfo  {journal} {Classical and Quantum Gravity}\ }\textbf
  {\bibinfo {volume} {25}},\ \bibinfo {pages} {184011} (\bibinfo {year}
  {2008}{\natexlab{a}})},\ \Eprint {https://arxiv.org/abs/0805.1689}
  {arXiv:0805.1689 [gr-qc]} \BibitemShut {NoStop}%
\bibitem [{\citenamefont {Zackay}\ \emph {et~al.}(2018)\citenamefont {Zackay},
  \citenamefont {Dai},\ and\ \citenamefont {Venumadhav}}]{Zackay_2018}%
  \BibitemOpen
  \bibfield  {author} {\bibinfo {author} {\bibfnamefont {B.}~\bibnamefont
  {Zackay}}, \bibinfo {author} {\bibfnamefont {L.}~\bibnamefont {Dai}},\ and\
  \bibinfo {author} {\bibfnamefont {T.}~\bibnamefont {Venumadhav}},\
  }\href@noop {} {\bibinfo {title} {Relative binning and fast likelihood
  evaluation for gravitational wave parameter estimation}} (\bibinfo {year}
  {2018}),\ \Eprint {https://arxiv.org/abs/1806.08792} {arXiv:1806.08792
  [astro-ph.IM]} \BibitemShut {NoStop}%
\bibitem [{\citenamefont {Smith}\ \emph {et~al.}(2016)\citenamefont {Smith}
  \emph {et~al.}}]{Smith_2016}%
  \BibitemOpen
  \bibfield  {author} {\bibinfo {author} {\bibfnamefont {R.}~\bibnamefont
  {Smith}} \emph {et~al.},\ }\bibfield  {title} {\bibinfo {title} {Fast and
  accurate inference on gravitational waves from precessing compact binaries},\
  }\bibfield  {journal} {\bibinfo  {journal} {Physical Review D}\ }\textbf
  {\bibinfo {volume} {94}},\ \href {https://doi.org/10.1103/physrevd.94.044031}
  {10.1103/physrevd.94.044031} (\bibinfo {year} {2016}),\ \Eprint
  {https://arxiv.org/abs/1604.08253} {arXiv:1604.08253 [gr-qc]} \BibitemShut
  {NoStop}%
\bibitem [{\citenamefont {Qi}\ and\ \citenamefont {Raymond}(2020)}]{Qi_2020}%
  \BibitemOpen
  \bibfield  {author} {\bibinfo {author} {\bibfnamefont {H.}~\bibnamefont
  {Qi}}\ and\ \bibinfo {author} {\bibfnamefont {V.}~\bibnamefont {Raymond}},\
  }\href@noop {} {\bibinfo {title} {Pyroq: a python-based reduced order
  quadrature building code for fast gravitational wave inference}} (\bibinfo
  {year} {2020}),\ \Eprint {https://arxiv.org/abs/2009.13812} {arXiv:2009.13812
  [gr-qc]} \BibitemShut {NoStop}%
\bibitem [{\citenamefont {Morisaki}\ and\ \citenamefont
  {Raymond}(2020)}]{Morisaki_2020}%
  \BibitemOpen
  \bibfield  {author} {\bibinfo {author} {\bibfnamefont {S.}~\bibnamefont
  {Morisaki}}\ and\ \bibinfo {author} {\bibfnamefont {V.}~\bibnamefont
  {Raymond}},\ }\bibfield  {title} {\bibinfo {title} {Rapid parameter
  estimation of gravitational waves from binary neutron star coalescence using
  focused reduced order quadrature},\ }\bibfield  {journal} {\bibinfo
  {journal} {Physical Review D}\ }\textbf {\bibinfo {volume} {102}},\ \href
  {https://doi.org/10.1103/physrevd.102.104020} {10.1103/physrevd.102.104020}
  (\bibinfo {year} {2020}),\ \Eprint {https://arxiv.org/abs/2007.09108}
  {arXiv:2007.09108 [gr-qc]} \BibitemShut {NoStop}%
\bibitem [{\citenamefont {Cornish}(2021)}]{Cornish_2021_RapidPE}%
  \BibitemOpen
  \bibfield  {author} {\bibinfo {author} {\bibfnamefont {N.~J.}\ \bibnamefont
  {Cornish}},\ }\href@noop {} {\bibinfo {title} {Rapid and robust parameter
  inference for binary mergers}} (\bibinfo {year} {2021}),\ \Eprint
  {https://arxiv.org/abs/2101.01188} {arXiv:2101.01188 [gr-qc]} \BibitemShut
  {NoStop}%
\bibitem [{\citenamefont {Ashton}\ \emph {et~al.}(2019)\citenamefont {Ashton}
  \emph {et~al.}}]{Ashton_2019}%
  \BibitemOpen
  \bibfield  {author} {\bibinfo {author} {\bibfnamefont {G.}~\bibnamefont
  {Ashton}} \emph {et~al.},\ }\bibfield  {title} {\bibinfo {title} {Bilby: A
  user-friendly bayesian inference library for gravitational-wave astronomy},\
  }\href {https://doi.org/10.3847/1538-4365/ab06fc} {\bibfield  {journal}
  {\bibinfo  {journal} {The Astrophysical Journal Supplement Series}\ }\textbf
  {\bibinfo {volume} {241}},\ \bibinfo {pages} {27} (\bibinfo {year} {2019})},\
  \Eprint {https://arxiv.org/abs/1811.02042} {arXiv:1811.02042 [astro-ph.IM]}
  \BibitemShut {NoStop}%
\bibitem [{\citenamefont {Romero-Shaw}\ \emph {et~al.}(2020)\citenamefont
  {Romero-Shaw} \emph {et~al.}}]{Romero_Shaw_2020}%
  \BibitemOpen
  \bibfield  {author} {\bibinfo {author} {\bibfnamefont {I.~M.}\ \bibnamefont
  {Romero-Shaw}} \emph {et~al.},\ }\bibfield  {title} {\bibinfo {title}
  {Bayesian inference for compact binary coalescences with bilby: Validation
  and application to the first ligo--virgo gravitational-wave transient
  catalogue},\ }\href {https://doi.org/10.1093/mnras/staa2850} {\bibfield
  {journal} {\bibinfo  {journal} {Monthly Notices of the Royal Astronomical
  Society}\ }\textbf {\bibinfo {volume} {499}},\ \bibinfo {pages} {3295}
  (\bibinfo {year} {2020})},\ \Eprint {https://arxiv.org/abs/2006.00714}
  {arXiv:2006.00714 [astro-ph.IM]} \BibitemShut {NoStop}%
\bibitem [{\citenamefont {Biwer}\ \emph {et~al.}(2019)\citenamefont {Biwer}
  \emph {et~al.}}]{Biwer_2019}%
  \BibitemOpen
  \bibfield  {author} {\bibinfo {author} {\bibfnamefont {C.~M.}\ \bibnamefont
  {Biwer}} \emph {et~al.},\ }\bibfield  {title} {\bibinfo {title} {Pycbc
  inference: A python-based parameter estimation toolkit for compact binary
  coalescence signals},\ }\href {https://doi.org/10.1088/1538-3873/aaef0b}
  {\bibfield  {journal} {\bibinfo  {journal} {Publications of the Astronomical
  Society of the Pacific}\ }\textbf {\bibinfo {volume} {131}},\ \bibinfo
  {pages} {024503} (\bibinfo {year} {2019})},\ \Eprint
  {https://arxiv.org/abs/1807.10312} {arXiv:1807.10312 [astro-ph.IM]}
  \BibitemShut {NoStop}%
\bibitem [{\citenamefont {Aasi}\ \emph {et~al.}(2013)\citenamefont {Aasi} \emph
  {et~al.}}]{Aasi_2013}%
  \BibitemOpen
  \bibfield  {author} {\bibinfo {author} {\bibfnamefont {J.}~\bibnamefont
  {Aasi}} \emph {et~al.} (\bibinfo {collaboration} {The LIGO Scientific
  Collaboration and Virgo Collaboration}),\ }\bibfield  {title} {\bibinfo
  {title} {Parameter estimation for compact binary coalescence signals with the
  first generation gravitational-wave detector network},\ }\bibfield  {journal}
  {\bibinfo  {journal} {Physical Review D}\ }\textbf {\bibinfo {volume} {88}},\
  \href {https://doi.org/10.1103/physrevd.88.062001}
  {10.1103/physrevd.88.062001} (\bibinfo {year} {2013}),\ \Eprint
  {https://arxiv.org/abs/1304.1775} {arXiv:1304.1775 [gr-qc]} \BibitemShut
  {NoStop}%
\bibitem [{\citenamefont {Jaranowski}\ and\ \citenamefont
  {Kr{\'o}lak}(2005)}]{Jaranowski_2005}%
  \BibitemOpen
  \bibfield  {author} {\bibinfo {author} {\bibfnamefont {P.}~\bibnamefont
  {Jaranowski}}\ and\ \bibinfo {author} {\bibfnamefont {A.}~\bibnamefont
  {Kr{\'o}lak}},\ }\bibfield  {title} {\bibinfo {title} {Gravitational-wave
  data analysis. formalism and sample applications: The gaussian case},\
  }\href@noop {} {\bibfield  {journal} {\bibinfo  {journal} {Living Reviews in
  Relativity}\ }\textbf {\bibinfo {volume} {8}} (\bibinfo {year} {2005})},\
  \Eprint {https://arxiv.org/abs/0711.1115} {arXiv:0711.1115 [gr-qc]}
  \BibitemShut {NoStop}%
\bibitem [{\citenamefont {Poisson}\ and\ \citenamefont
  {Will}(1995)}]{Poisson_1995}%
  \BibitemOpen
  \bibfield  {author} {\bibinfo {author} {\bibfnamefont {E.}~\bibnamefont
  {Poisson}}\ and\ \bibinfo {author} {\bibfnamefont {C.~M.}\ \bibnamefont
  {Will}},\ }\bibfield  {title} {\bibinfo {title} {Gravitational waves from
  inspiralling compact binaries: Parameter estimation using
  second-post-newtonian waveforms},\ }\href
  {https://doi.org/10.1103/physrevd.52.848} {\bibfield  {journal} {\bibinfo
  {journal} {Physical Review D}\ }\textbf {\bibinfo {volume} {52}},\ \bibinfo
  {pages} {848} (\bibinfo {year} {1995})},\ \Eprint
  {https://arxiv.org/abs/gr-qc/9502040} {arXiv:gr-qc/9502040 [gr-qc]}
  \BibitemShut {NoStop}%
\bibitem [{\citenamefont {Perraudin}\ and\ \citenamefont
  {Vandergheynst}(2017)}]{Perraudin_2017}%
  \BibitemOpen
  \bibfield  {author} {\bibinfo {author} {\bibfnamefont {N.}~\bibnamefont
  {Perraudin}}\ and\ \bibinfo {author} {\bibfnamefont {P.}~\bibnamefont
  {Vandergheynst}},\ }\bibfield  {title} {\bibinfo {title} {Stationary signal
  processing on graphs},\ }\href {https://doi.org/10.1109/tsp.2017.2690388}
  {\bibfield  {journal} {\bibinfo  {journal} {IEEE Transactions on Signal
  Processing}\ }\textbf {\bibinfo {volume} {65}},\ \bibinfo {pages} {3462}
  (\bibinfo {year} {2017})}\BibitemShut {NoStop}%
\bibitem [{\citenamefont {Littenberg}\ and\ \citenamefont
  {Cornish}(2015)}]{Littenberg_2015}%
  \BibitemOpen
  \bibfield  {author} {\bibinfo {author} {\bibfnamefont {T.~B.}\ \bibnamefont
  {Littenberg}}\ and\ \bibinfo {author} {\bibfnamefont {N.~J.}\ \bibnamefont
  {Cornish}},\ }\bibfield  {title} {\bibinfo {title} {Bayesian inference for
  spectral estimation of gravitational wave detector noise},\ }\bibfield
  {journal} {\bibinfo  {journal} {Physical Review D}\ }\textbf {\bibinfo
  {volume} {91}},\ \href {https://doi.org/10.1103/physrevd.91.084034}
  {10.1103/physrevd.91.084034} (\bibinfo {year} {2015}),\ \Eprint
  {https://arxiv.org/abs/1410.3852} {arXiv:1410.3852 [gr-qc]} \BibitemShut
  {NoStop}%
\bibitem [{\citenamefont {van~der Sluys}\ \emph
  {et~al.}(2008{\natexlab{b}})\citenamefont {van~der Sluys} \emph
  {et~al.}}]{van_der_Sluys_2007}%
  \BibitemOpen
  \bibfield  {author} {\bibinfo {author} {\bibfnamefont {M.}~\bibnamefont
  {van~der Sluys}} \emph {et~al.},\ }\bibfield  {title} {\bibinfo {title}
  {Gravitational-wave astronomy with inspiral signals of spinning
  compact-object binaries},\ }\href {https://doi.org/10.1086/595279} {\bibfield
   {journal} {\bibinfo  {journal} {The Astrophysical Journal}\ }\textbf
  {\bibinfo {volume} {688}},\ \bibinfo {pages} {L61} (\bibinfo {year}
  {2008}{\natexlab{b}})},\ \Eprint {https://arxiv.org/abs/0710.1897}
  {arXiv:0710.1897 [astro-ph]} \BibitemShut {NoStop}%
\bibitem [{\citenamefont {Blackburn}\ \emph {et~al.}(2008)\citenamefont
  {Blackburn} \emph {et~al.}}]{Blackburn_2008}%
  \BibitemOpen
  \bibfield  {author} {\bibinfo {author} {\bibfnamefont {L.}~\bibnamefont
  {Blackburn}} \emph {et~al.},\ }\bibfield  {title} {\bibinfo {title} {The lsc
  glitch group: Monitoring noise transients during the fifth ligo science
  run},\ }\href {https://doi.org/10.1088/0264-9381/25/18/184004} {\bibfield
  {journal} {\bibinfo  {journal} {Classical and Quantum Gravity}\ }\textbf
  {\bibinfo {volume} {25}},\ \bibinfo {pages} {184004} (\bibinfo {year}
  {2008})},\ \Eprint {https://arxiv.org/abs/0804.0800} {arXiv:0804.0800
  [gr-qc]} \BibitemShut {NoStop}%
\bibitem [{\citenamefont {Abbott}\ \emph
  {et~al.}(2016{\natexlab{b}})\citenamefont {Abbott} \emph
  {et~al.}}]{Abbott_2016_TransientCharacterization}%
  \BibitemOpen
  \bibfield  {author} {\bibinfo {author} {\bibfnamefont {B.~P.}\ \bibnamefont
  {Abbott}} \emph {et~al.} (\bibinfo {collaboration} {The LIGO Scientific
  Collaboration and Virgo Collaboration}),\ }\bibfield  {title} {\bibinfo
  {title} {Characterization of transient noise in advanced ligo relevant to
  gravitational wave signal gw150914},\ }\href
  {https://doi.org/10.1088/0264-9381/33/13/134001} {\bibfield  {journal}
  {\bibinfo  {journal} {Classical and Quantum Gravity}\ }\textbf {\bibinfo
  {volume} {33}},\ \bibinfo {pages} {134001} (\bibinfo {year}
  {2016}{\natexlab{b}})},\ \Eprint {https://arxiv.org/abs/1602.03844}
  {arXiv:1602.03844 [gr-qc]} \BibitemShut {NoStop}%
\bibitem [{\citenamefont {Abbott}\ \emph
  {et~al.}(2019{\natexlab{a}})\citenamefont {Abbott} \emph
  {et~al.}}]{Abbott_2019}%
  \BibitemOpen
  \bibfield  {author} {\bibinfo {author} {\bibfnamefont {R.}~\bibnamefont
  {Abbott}} \emph {et~al.} (\bibinfo {collaboration} {The LIGO Scientific
  Collaboration and the Virgo Collaboration}),\ }\href@noop {} {\bibinfo
  {title} {Open data from the first and second observing runs of advanced ligo
  and advanced virgo}} (\bibinfo {year} {2019}{\natexlab{a}}),\ \Eprint
  {https://arxiv.org/abs/1912.11716} {arXiv:1912.11716 [gr-qc]} \BibitemShut
  {NoStop}%
\bibitem [{\citenamefont {Abbott}\ \emph
  {et~al.}(2020{\natexlab{a}})\citenamefont {Abbott} \emph
  {et~al.}}]{Abbott_2020}%
  \BibitemOpen
  \bibfield  {author} {\bibinfo {author} {\bibfnamefont {B.~P.}\ \bibnamefont
  {Abbott}} \emph {et~al.} (\bibinfo {collaboration} {The LIGO Scientific
  Collaboration and Virgo Collaboration}),\ }\bibfield  {title} {\bibinfo
  {title} {A guide to ligo-virgo detector noise and extraction of transient
  gravitational-wave signals},\ }\href
  {https://doi.org/10.1088/1361-6382/ab685e} {\bibfield  {journal} {\bibinfo
  {journal} {Classical and Quantum Gravity}\ }\textbf {\bibinfo {volume}
  {37}},\ \bibinfo {pages} {055002} (\bibinfo {year} {2020}{\natexlab{a}})},\
  \Eprint {https://arxiv.org/abs/1908.11170} {arXiv:1908.11170 [gr-qc]}
  \BibitemShut {NoStop}%
\bibitem [{\citenamefont {Chatziioannou}\ \emph {et~al.}(2019)\citenamefont
  {Chatziioannou} \emph {et~al.}}]{Chatziioannou_2019}%
  \BibitemOpen
  \bibfield  {author} {\bibinfo {author} {\bibfnamefont {K.}~\bibnamefont
  {Chatziioannou}} \emph {et~al.},\ }\bibfield  {title} {\bibinfo {title}
  {Noise spectral estimation methods and their impact on gravitational wave
  measurement of compact binary mergers},\ }\bibfield  {journal} {\bibinfo
  {journal} {Physical Review D}\ }\textbf {\bibinfo {volume} {100}},\ \href
  {https://doi.org/10.1103/physrevd.100.104004} {10.1103/physrevd.100.104004}
  (\bibinfo {year} {2019}),\ \Eprint {https://arxiv.org/abs/1907.06540}
  {arXiv:1907.06540 [gr-qc]} \BibitemShut {NoStop}%
\bibitem [{\citenamefont {Cornish}(2020)}]{Cornish_2020}%
  \BibitemOpen
  \bibfield  {author} {\bibinfo {author} {\bibfnamefont {N.~J.}\ \bibnamefont
  {Cornish}},\ }\bibfield  {title} {\bibinfo {title} {Time-frequency analysis
  of gravitational wave data},\ }\href
  {https://doi.org/10.1103/PhysRevD.102.124038} {\bibfield  {journal} {\bibinfo
   {journal} {Physical Review D}\ }\textbf {\bibinfo {volume} {102}},\ \bibinfo
  {pages} {124038} (\bibinfo {year} {2020})},\ \Eprint
  {https://arxiv.org/abs/2009.00043} {arXiv:2009.00043 [gr-qc]} \BibitemShut
  {NoStop}%
\bibitem [{\citenamefont {Cornish}\ \emph {et~al.}(2021)\citenamefont {Cornish}
  \emph {et~al.}}]{Cornish_2021_BayesWave}%
  \BibitemOpen
  \bibfield  {author} {\bibinfo {author} {\bibfnamefont {N.~J.}\ \bibnamefont
  {Cornish}} \emph {et~al.},\ }\bibfield  {title} {\bibinfo {title} {Bayeswave
  analysis pipeline in the era of gravitational wave observations},\ }\bibfield
   {journal} {\bibinfo  {journal} {Physical Review D}\ }\textbf {\bibinfo
  {volume} {103}},\ \href {https://doi.org/10.1103/physrevd.103.044006}
  {10.1103/physrevd.103.044006} (\bibinfo {year} {2021}),\ \Eprint
  {https://arxiv.org/abs/2011.09494} {arXiv:2011.09494 [gr-qc]} \BibitemShut
  {NoStop}%
\bibitem [{\citenamefont {Chatziioannou}\ \emph {et~al.}(2021)\citenamefont
  {Chatziioannou} \emph {et~al.}}]{Chatziioannou_2021}%
  \BibitemOpen
  \bibfield  {author} {\bibinfo {author} {\bibfnamefont {K.}~\bibnamefont
  {Chatziioannou}} \emph {et~al.},\ }\bibfield  {title} {\bibinfo {title}
  {Modeling compact binary signals and instrumental glitches in gravitational
  wave data},\ }\bibfield  {journal} {\bibinfo  {journal} {Physical Review D}\
  }\textbf {\bibinfo {volume} {103}},\ \href
  {https://doi.org/10.1103/physrevd.103.044013} {10.1103/physrevd.103.044013}
  (\bibinfo {year} {2021}),\ \Eprint {https://arxiv.org/abs/2101.01200}
  {arXiv:2101.01200 [gr-qc]} \BibitemShut {NoStop}%
\bibitem [{\citenamefont {Mozzon}\ \emph {et~al.}(2020)\citenamefont {Mozzon}
  \emph {et~al.}}]{Mozzon_2020}%
  \BibitemOpen
  \bibfield  {author} {\bibinfo {author} {\bibfnamefont {S.}~\bibnamefont
  {Mozzon}} \emph {et~al.},\ }\bibfield  {title} {\bibinfo {title} {Dynamic
  normalization for compact binary coalescence searches in non-stationary
  noise},\ }\href {https://doi.org/10.1088/1361-6382/abac6c} {\bibfield
  {journal} {\bibinfo  {journal} {Classical and Quantum Gravity}\ }\textbf
  {\bibinfo {volume} {37}},\ \bibinfo {pages} {215014} (\bibinfo {year}
  {2020})},\ \Eprint {https://arxiv.org/abs/2002.09407} {arXiv:2002.09407
  [astro-ph.IM]} \BibitemShut {NoStop}%
\bibitem [{\citenamefont {Biscoveanu}\ \emph {et~al.}(2020)\citenamefont
  {Biscoveanu} \emph {et~al.}}]{Biscoveanu_2020}%
  \BibitemOpen
  \bibfield  {author} {\bibinfo {author} {\bibfnamefont {S.}~\bibnamefont
  {Biscoveanu}} \emph {et~al.},\ }\bibfield  {title} {\bibinfo {title}
  {Quantifying the effect of power spectral density uncertainty on
  gravitational-wave parameter estimation for compact binary sources},\
  }\bibfield  {journal} {\bibinfo  {journal} {Physical Review D}\ }\textbf
  {\bibinfo {volume} {102}},\ \href
  {https://doi.org/10.1103/physrevd.102.023008} {10.1103/physrevd.102.023008}
  (\bibinfo {year} {2020}),\ \Eprint {https://arxiv.org/abs/2004.05149}
  {arXiv:2004.05149 [astro-ph.HE]} \BibitemShut {NoStop}%
\bibitem [{\citenamefont {R{\"o}ver}(2011)}]{Rover_2011}%
  \BibitemOpen
  \bibfield  {author} {\bibinfo {author} {\bibfnamefont {C.}~\bibnamefont
  {R{\"o}ver}},\ }\bibfield  {title} {\bibinfo {title} {Student-t based filter
  for robust signal detection},\ }\bibfield  {journal} {\bibinfo  {journal}
  {Physical Review D}\ }\textbf {\bibinfo {volume} {84}},\ \href
  {https://doi.org/10.1103/physrevd.84.122004} {10.1103/physrevd.84.122004}
  (\bibinfo {year} {2011}),\ \Eprint {https://arxiv.org/abs/1109.0442}
  {arXiv:1109.0442 [physics.data-an]} \BibitemShut {NoStop}%
\bibitem [{\citenamefont {Wainstein}\ and\ \citenamefont
  {Zubakov}(1962)}]{Wainstein_1962}%
  \BibitemOpen
  \bibfield  {author} {\bibinfo {author} {\bibfnamefont {L.~A.}\ \bibnamefont
  {Wainstein}}\ and\ \bibinfo {author} {\bibfnamefont {V.~D.}\ \bibnamefont
  {Zubakov}},\ }\href@noop {} {\emph {\bibinfo {title} {Extraction of Signals
  from Noise}}}\ (\bibinfo  {publisher} {Prentice-Hall International},\
  \bibinfo {year} {1962})\BibitemShut {NoStop}%
\bibitem [{\citenamefont {Trotta}(2008)}]{Trotta_2008}%
  \BibitemOpen
  \bibfield  {author} {\bibinfo {author} {\bibfnamefont {R.}~\bibnamefont
  {Trotta}},\ }\bibfield  {title} {\bibinfo {title} {Bayes in the sky: Bayesian
  inference and model selection in cosmology},\ }\href
  {https://doi.org/10.1080/00107510802066753} {\bibfield  {journal} {\bibinfo
  {journal} {Contemporary Physics}\ }\textbf {\bibinfo {volume} {49}},\
  \bibinfo {pages} {71} (\bibinfo {year} {2008})},\ \Eprint
  {https://arxiv.org/abs/0803.4089} {arXiv:0803.4089 [astro-ph]} \BibitemShut
  {NoStop}%
\bibitem [{\citenamefont {Robert}\ and\ \citenamefont
  {Rousseau}(2010)}]{Robert_2010}%
  \BibitemOpen
  \bibfield  {author} {\bibinfo {author} {\bibfnamefont {C.~P.}\ \bibnamefont
  {Robert}}\ and\ \bibinfo {author} {\bibfnamefont {J.}~\bibnamefont
  {Rousseau}},\ }\href@noop {} {\bibinfo {title} {On bayesian data analysis}}
  (\bibinfo {year} {2010}),\ \Eprint {https://arxiv.org/abs/1001.4656}
  {arXiv:1001.4656 [stat.ME]} \BibitemShut {NoStop}%
\bibitem [{\citenamefont {Vitale}\ \emph {et~al.}(2017)\citenamefont {Vitale}
  \emph {et~al.}}]{Vitale_2017}%
  \BibitemOpen
  \bibfield  {author} {\bibinfo {author} {\bibfnamefont {S.}~\bibnamefont
  {Vitale}} \emph {et~al.},\ }\bibfield  {title} {\bibinfo {title} {Impact of
  bayesian priors on the characterization of binary black hole coalescences},\
  }\bibfield  {journal} {\bibinfo  {journal} {Physical Review Letters}\
  }\textbf {\bibinfo {volume} {119}},\ \href
  {https://doi.org/10.1103/physrevlett.119.251103}
  {10.1103/physrevlett.119.251103} (\bibinfo {year} {2017}),\ \Eprint
  {https://arxiv.org/abs/1707.04637} {arXiv:1707.04637 [gr-qc]} \BibitemShut
  {NoStop}%
\bibitem [{\citenamefont {Gelman}\ and\ \citenamefont
  {Meng}(1998)}]{Gelman_1998}%
  \BibitemOpen
  \bibfield  {author} {\bibinfo {author} {\bibfnamefont {A.}~\bibnamefont
  {Gelman}}\ and\ \bibinfo {author} {\bibfnamefont {X.-L.}\ \bibnamefont
  {Meng}},\ }\bibfield  {title} {\bibinfo {title} {Simulating normalizing
  constants: From importance sampling to bridge sampling to path sampling},\
  }\href {http://www.jstor.org/stable/2676756} {\bibfield  {journal} {\bibinfo
  {journal} {Statistical Science}\ }\textbf {\bibinfo {volume} {13}},\ \bibinfo
  {pages} {163} (\bibinfo {year} {1998})}\BibitemShut {NoStop}%
\bibitem [{\citenamefont {Hogg}\ and\ \citenamefont
  {Foreman-Mackey}(2018)}]{Hogg_2018}%
  \BibitemOpen
  \bibfield  {author} {\bibinfo {author} {\bibfnamefont {D.~W.}\ \bibnamefont
  {Hogg}}\ and\ \bibinfo {author} {\bibfnamefont {D.}~\bibnamefont
  {Foreman-Mackey}},\ }\bibfield  {title} {\bibinfo {title} {Data analysis
  recipes: Using markov chain monte carlo},\ }\href
  {https://doi.org/10.3847/1538-4365/aab76e} {\bibfield  {journal} {\bibinfo
  {journal} {The Astrophysical Journal Supplement Series}\ }\textbf {\bibinfo
  {volume} {236}},\ \bibinfo {pages} {11} (\bibinfo {year} {2018})},\ \Eprint
  {https://arxiv.org/abs/1710.06068} {arXiv:1710.06068 [astro-ph.IM]}
  \BibitemShut {NoStop}%
\bibitem [{\citenamefont {Christensen}\ and\ \citenamefont
  {Meyer}(2001)}]{Christensen_2001}%
  \BibitemOpen
  \bibfield  {author} {\bibinfo {author} {\bibfnamefont {N.}~\bibnamefont
  {Christensen}}\ and\ \bibinfo {author} {\bibfnamefont {R.}~\bibnamefont
  {Meyer}},\ }\bibfield  {title} {\bibinfo {title} {Using markov chain monte
  carlo methods for estimating parameters with gravitational radiation data},\
  }\bibfield  {journal} {\bibinfo  {journal} {Physical Review D}\ }\textbf
  {\bibinfo {volume} {64}},\ \href {https://doi.org/10.1103/physrevd.64.022001}
  {10.1103/physrevd.64.022001} (\bibinfo {year} {2001}),\ \Eprint
  {https://arxiv.org/abs/gr-qc/0102018} {arXiv:gr-qc/0102018 [gr-qc]}
  \BibitemShut {NoStop}%
\bibitem [{\citenamefont {Speagle}(2020)}]{Speagle_2020}%
  \BibitemOpen
  \bibfield  {author} {\bibinfo {author} {\bibfnamefont {J.~S.}\ \bibnamefont
  {Speagle}},\ }\bibfield  {title} {\bibinfo {title} {dynesty: a dynamic nested
  sampling package for estimating bayesian posteriors and evidences},\ }\href
  {https://doi.org/10.1093/mnras/staa278} {\bibfield  {journal} {\bibinfo
  {journal} {Monthly Notices of the Royal Astronomical Society}\ }\textbf
  {\bibinfo {volume} {493}},\ \bibinfo {pages} {3132} (\bibinfo {year}
  {2020})},\ \Eprint {https://arxiv.org/abs/1904.02180} {arXiv:1904.02180
  [astro-ph.IM]} \BibitemShut {NoStop}%
\bibitem [{\citenamefont {Scargle}(2016)}]{Scargle_2016}%
  \BibitemOpen
  \bibfield  {author} {\bibinfo {author} {\bibfnamefont {J.~D.}\ \bibnamefont
  {Scargle}},\ }\href@noop {} {\bibinfo {title} {Degeneracy of arma time series
  models and the arrow of time}} (\bibinfo {year} {2016})\BibitemShut {NoStop}%
\bibitem [{\citenamefont {Vajente}\ \emph {et~al.}(2020)\citenamefont {Vajente}
  \emph {et~al.}}]{Vajente_2020}%
  \BibitemOpen
  \bibfield  {author} {\bibinfo {author} {\bibfnamefont {G.}~\bibnamefont
  {Vajente}} \emph {et~al.},\ }\bibfield  {title} {\bibinfo {title}
  {Machine-learning nonstationary noise out of gravitational-wave detectors},\
  }\bibfield  {journal} {\bibinfo  {journal} {Physical Review D}\ }\textbf
  {\bibinfo {volume} {101}},\ \href
  {https://doi.org/10.1103/physrevd.101.042003} {10.1103/physrevd.101.042003}
  (\bibinfo {year} {2020}),\ \Eprint {https://arxiv.org/abs/1911.09083}
  {arXiv:1911.09083 [gr-qc]} \BibitemShut {NoStop}%
\bibitem [{\citenamefont {Cram{\'e}r}(1961)}]{Cramer_1961}%
  \BibitemOpen
  \bibfield  {author} {\bibinfo {author} {\bibfnamefont {H.}~\bibnamefont
  {Cram{\'e}r}},\ }\bibfield  {title} {\bibinfo {title} {On some classes of
  nonstationary stochastic processes},\ }in\ \href
  {https://projecteuclid.org/euclid.bsmsp/1200512593} {\emph {\bibinfo
  {booktitle} {Proceedings of the Fourth Berkeley Symposium on Mathematical
  Statistics and Probability, Volume 2: Contributions to Probability Theory}}}\
  (\bibinfo  {publisher} {University of California Press},\ \bibinfo {address}
  {Berkeley, California},\ \bibinfo {year} {1961})\ pp.\ \bibinfo {pages}
  {57--78}\BibitemShut {NoStop}%
\bibitem [{\citenamefont {Hebbal}\ \emph {et~al.}(2019)\citenamefont {Hebbal}
  \emph {et~al.}}]{Hebbal_2019}%
  \BibitemOpen
  \bibfield  {author} {\bibinfo {author} {\bibfnamefont {A.}~\bibnamefont
  {Hebbal}} \emph {et~al.},\ }\href@noop {} {\bibinfo {title} {Bayesian
  optimization using deep gaussian processes}} (\bibinfo {year} {2019}),\
  \Eprint {https://arxiv.org/abs/1905.03350} {arXiv:1905.03350 [stat.ML]}
  \BibitemShut {NoStop}%
\bibitem [{\citenamefont {Kamen}\ and\ \citenamefont
  {Sills}(1993)}]{Kamen_1993}%
  \BibitemOpen
  \bibfield  {author} {\bibinfo {author} {\bibfnamefont {E.~W.}\ \bibnamefont
  {Kamen}}\ and\ \bibinfo {author} {\bibfnamefont {J.~A.}\ \bibnamefont
  {Sills}},\ }\bibfield  {title} {\bibinfo {title} {The frequency response
  function of a linear time-varying system},\ }\href
  {https://doi.org/https://doi.org/10.1016/S1474-6670(17)49134-8} {\bibfield
  {journal} {\bibinfo  {journal} {IFAC Proceedings Volumes}\ }\textbf {\bibinfo
  {volume} {26}},\ \bibinfo {pages} {315} (\bibinfo {year} {1993})},\ \bibinfo
  {note} {12th Triennal Wold Congress of the International Federation of
  Automatic control. Volume 1 Theory, Sydney, Australia, 18-23
  July}\BibitemShut {NoStop}%
\bibitem [{\citenamefont {Loynes}(1968)}]{Loynes_1968}%
  \BibitemOpen
  \bibfield  {author} {\bibinfo {author} {\bibfnamefont {R.~M.}\ \bibnamefont
  {Loynes}},\ }\bibfield  {title} {\bibinfo {title} {On the concept of the
  spectrum for non-stationary processes},\ }\href
  {http://www.jstor.org/stable/2984457} {\bibfield  {journal} {\bibinfo
  {journal} {Journal of the Royal Statistical Society. Series B
  (Methodological)}\ }\textbf {\bibinfo {volume} {30}},\ \bibinfo {pages} {1}
  (\bibinfo {year} {1968})}\BibitemShut {NoStop}%
\bibitem [{\citenamefont {Huang}\ \emph {et~al.}(1998)\citenamefont {Huang}
  \emph {et~al.}}]{Huang_1998}%
  \BibitemOpen
  \bibfield  {author} {\bibinfo {author} {\bibfnamefont {N.}~\bibnamefont
  {Huang}} \emph {et~al.},\ }\bibfield  {title} {\bibinfo {title} {The
  empirical mode decomposition and the hilbert spectrum for nonlinear and
  non-stationary time series analysis},\ }\href
  {https://doi.org/10.1098/rspa.1998.0193} {\bibfield  {journal} {\bibinfo
  {journal} {Proceedings of the Royal Society of London. Series A:
  Mathematical, Physical and Engineering Sciences}\ }\textbf {\bibinfo {volume}
  {454}},\ \bibinfo {pages} {903} (\bibinfo {year} {1998})}\BibitemShut
  {NoStop}%
\bibitem [{\citenamefont {Sun}\ \emph {et~al.}(2020)\citenamefont {Sun} \emph
  {et~al.}}]{Sun_2020}%
  \BibitemOpen
  \bibfield  {author} {\bibinfo {author} {\bibfnamefont {L.}~\bibnamefont
  {Sun}} \emph {et~al.},\ }\bibfield  {title} {\bibinfo {title}
  {Characterization of systematic error in advanced ligo calibration},\ }\href
  {https://doi.org/10.1088/1361-6382/abb14e} {\bibfield  {journal} {\bibinfo
  {journal} {Classical and Quantum Gravity}\ }\textbf {\bibinfo {volume}
  {37}},\ \bibinfo {pages} {225008} (\bibinfo {year} {2020})},\ \Eprint
  {https://arxiv.org/abs/2005.02531} {arXiv:2005.02531 [astro-ph.IM]}
  \BibitemShut {NoStop}%
\bibitem [{\citenamefont {Vitale}\ \emph {et~al.}(2020)\citenamefont {Vitale}
  \emph {et~al.}}]{Vitale_2020}%
  \BibitemOpen
  \bibfield  {author} {\bibinfo {author} {\bibfnamefont {S.}~\bibnamefont
  {Vitale}} \emph {et~al.},\ }\href@noop {} {\bibinfo {title} {physical: A
  physical approach to the marginalization of ligo calibration uncertainties}}
  (\bibinfo {year} {2020}),\ \Eprint {https://arxiv.org/abs/2009.10192}
  {arXiv:2009.10192 [gr-qc]} \BibitemShut {NoStop}%
\bibitem [{\citenamefont {Payne}\ \emph {et~al.}(2020)\citenamefont {Payne}
  \emph {et~al.}}]{Payne_2020}%
  \BibitemOpen
  \bibfield  {author} {\bibinfo {author} {\bibfnamefont {E.}~\bibnamefont
  {Payne}} \emph {et~al.},\ }\bibfield  {title} {\bibinfo {title}
  {Gravitational-wave astronomy with a physical calibration model},\ }\bibfield
   {journal} {\bibinfo  {journal} {Physical Review D}\ }\textbf {\bibinfo
  {volume} {102}},\ \href {https://doi.org/10.1103/physrevd.102.122004}
  {10.1103/physrevd.102.122004} (\bibinfo {year} {2020}),\ \Eprint
  {https://arxiv.org/abs/2009.10193} {arXiv:2009.10193 [astro-ph.IM]}
  \BibitemShut {NoStop}%
\bibitem [{\citenamefont {Abbott}\ \emph
  {et~al.}(2019{\natexlab{b}})\citenamefont {Abbott} \emph {et~al.}}]{GWTC-1}%
  \BibitemOpen
  \bibfield  {author} {\bibinfo {author} {\bibfnamefont {B.~P.}\ \bibnamefont
  {Abbott}} \emph {et~al.} (\bibinfo {collaboration} {The LIGO Scientific
  Collaboration and Virgo Collaboration}),\ }\bibfield  {title} {\bibinfo
  {title} {Gwtc-1: A gravitational-wave transient catalog of compact binary
  mergers observed by ligo and virgo during the first and second observing
  runs},\ }\bibfield  {journal} {\bibinfo  {journal} {Physical Review X}\
  }\textbf {\bibinfo {volume} {9}},\ \href
  {https://doi.org/10.1103/physrevx.9.031040} {10.1103/physrevx.9.031040}
  (\bibinfo {year} {2019}{\natexlab{b}}),\ \Eprint
  {https://arxiv.org/abs/1811.12907} {arXiv:1811.12907 [astro-ph.HE]}
  \BibitemShut {NoStop}%
\bibitem [{\citenamefont {Abbott}\ \emph
  {et~al.}(2019{\natexlab{c}})\citenamefont {Abbott} \emph {et~al.}}]{LIGO_O2}%
  \BibitemOpen
  \bibfield  {author} {\bibinfo {author} {\bibfnamefont {R.}~\bibnamefont
  {Abbott}} \emph {et~al.} (\bibinfo {collaboration} {The LIGO Scientific
  Collaboration and Virgo Collaboration}),\ }\href
  {https://doi.org/10.7935/CA75-FM95} {\bibinfo {title} {The o2 data release}}
  (\bibinfo {year} {2019}{\natexlab{c}})\BibitemShut {NoStop}%
\bibitem [{\citenamefont {Abbott}\ \emph
  {et~al.}(2020{\natexlab{b}})\citenamefont {Abbott} \emph {et~al.}}]{GWTC-2}%
  \BibitemOpen
  \bibfield  {author} {\bibinfo {author} {\bibfnamefont {R.}~\bibnamefont
  {Abbott}} \emph {et~al.} (\bibinfo {collaboration} {The LIGO Scientific
  Collaboration and Virgo Collaboration}),\ }\bibfield  {title} {\bibinfo
  {title} {Gwtc-2: Compact binary coalescences observed by ligo and virgo
  during the first half of the third observing run},\ }\Eprint
  {https://arxiv.org/abs/2010.14527} {arXiv:2010.14527 [gr-qc]}  (\bibinfo
  {year} {2020}{\natexlab{b}})\BibitemShut {NoStop}%
\bibitem [{\citenamefont {Talbot}\ and\ \citenamefont
  {Thrane}(2020)}]{Talbot_2020}%
  \BibitemOpen
  \bibfield  {author} {\bibinfo {author} {\bibfnamefont {C.}~\bibnamefont
  {Talbot}}\ and\ \bibinfo {author} {\bibfnamefont {E.}~\bibnamefont
  {Thrane}},\ }\bibfield  {title} {\bibinfo {title} {Gravitational-wave
  astronomy with an uncertain noise power spectral density},\ }\href
  {https://doi.org/10.1103/PhysRevResearch.2.043298} {\bibfield  {journal}
  {\bibinfo  {journal} {Physical Review Research}\ }\textbf {\bibinfo {volume}
  {2}},\ \bibinfo {pages} {043298} (\bibinfo {year} {2020})},\ \Eprint
  {https://arxiv.org/abs/2006.05292} {arXiv:2006.05292 [astro-ph.IM]}
  \BibitemShut {NoStop}%
\bibitem [{\citenamefont {Zackay}\ \emph {et~al.}(2019)\citenamefont {Zackay}
  \emph {et~al.}}]{Zackay_2019}%
  \BibitemOpen
  \bibfield  {author} {\bibinfo {author} {\bibfnamefont {B.}~\bibnamefont
  {Zackay}} \emph {et~al.},\ }\href@noop {} {\bibinfo {title} {Detecting
  gravitational waves in data with non-gaussian noise}} (\bibinfo {year}
  {2019}),\ \Eprint {https://arxiv.org/abs/1908.05644} {arXiv:1908.05644
  [astro-ph.IM]} \BibitemShut {NoStop}%
\bibitem [{\citenamefont {Venumadhav}\ \emph {et~al.}(2019)\citenamefont
  {Venumadhav} \emph {et~al.}}]{Venumadhav_2019}%
  \BibitemOpen
  \bibfield  {author} {\bibinfo {author} {\bibfnamefont {T.}~\bibnamefont
  {Venumadhav}} \emph {et~al.},\ }\bibfield  {title} {\bibinfo {title} {New
  search pipeline for compact binary mergers: Results for binary black holes in
  the first observing run of advanced ligo},\ }\bibfield  {journal} {\bibinfo
  {journal} {Physical Review D}\ }\textbf {\bibinfo {volume} {100}},\ \href
  {https://doi.org/10.1103/physrevd.100.023011} {10.1103/physrevd.100.023011}
  (\bibinfo {year} {2019}),\ \Eprint {https://arxiv.org/abs/1902.10341}
  {arXiv:1902.10341 [astro-ph.IM]} \BibitemShut {NoStop}%
\bibitem [{\citenamefont {Edwards}\ \emph {et~al.}(2020)\citenamefont {Edwards}
  \emph {et~al.}}]{Edwards_2020}%
  \BibitemOpen
  \bibfield  {author} {\bibinfo {author} {\bibfnamefont {M.~C.}\ \bibnamefont
  {Edwards}} \emph {et~al.},\ }\bibfield  {title} {\bibinfo {title}
  {Identifying and addressing nonstationary lisa noise},\ }\bibfield  {journal}
  {\bibinfo  {journal} {Physical Review D}\ }\textbf {\bibinfo {volume}
  {102}},\ \href {https://doi.org/10.1103/physrevd.102.084062}
  {10.1103/physrevd.102.084062} (\bibinfo {year} {2020}),\ \Eprint
  {https://arxiv.org/abs/2004.07515} {arXiv:2004.07515 [gr-qc]} \BibitemShut
  {NoStop}%
\bibitem [{\citenamefont {Huang}\ \emph {et~al.}(2020)\citenamefont {Huang}
  \emph {et~al.}}]{Huang_2020}%
  \BibitemOpen
  \bibfield  {author} {\bibinfo {author} {\bibfnamefont {Y.}~\bibnamefont
  {Huang}} \emph {et~al.},\ }\bibfield  {title} {\bibinfo {title} {Source
  properties of the lowest signal-to-noise-ratio binary black hole
  detections},\ }\bibfield  {journal} {\bibinfo  {journal} {Physical Review D}\
  }\textbf {\bibinfo {volume} {102}},\ \href
  {https://doi.org/10.1103/physrevd.102.103024} {10.1103/physrevd.102.103024}
  (\bibinfo {year} {2020}),\ \Eprint {https://arxiv.org/abs/2003.04513}
  {arXiv:2003.04513 [gr-qc]} \BibitemShut {NoStop}%
\bibitem [{\citenamefont {Vallisneri}(2008)}]{Vallisneri_2008}%
  \BibitemOpen
  \bibfield  {author} {\bibinfo {author} {\bibfnamefont {M.}~\bibnamefont
  {Vallisneri}},\ }\bibfield  {title} {\bibinfo {title} {Use and abuse of the
  fisher information matrix in the assessment of gravitational-wave
  parameter-estimation prospects},\ }\href
  {https://doi.org/10.1103/physrevd.77.042001} {\bibfield  {journal} {\bibinfo
  {journal} {Physical Review D}\ }\textbf {\bibinfo {volume} {77}},\ \bibinfo
  {pages} {042001} (\bibinfo {year} {2008})}\BibitemShut {NoStop}%
\bibitem [{\citenamefont {Vallisneri}(2011)}]{Vallisneri_2011}%
  \BibitemOpen
  \bibfield  {author} {\bibinfo {author} {\bibfnamefont {M.}~\bibnamefont
  {Vallisneri}},\ }\bibfield  {title} {\bibinfo {title} {Beyond the
  fisher-matrix formalism: Exact sampling distributions of the
  maximum-likelihood estimator in gravitational-wave parameter estimation},\
  }\bibfield  {journal} {\bibinfo  {journal} {Physical Review Letters}\
  }\textbf {\bibinfo {volume} {107}},\ \href
  {https://doi.org/10.1103/physrevlett.107.191104}
  {10.1103/physrevlett.107.191104} (\bibinfo {year} {2011}),\ \Eprint
  {https://arxiv.org/abs/1108.1158} {arXiv:1108.1158 [gr-qc]} \BibitemShut
  {NoStop}%
\bibitem [{\citenamefont {Sathyaprakash}\ and\ \citenamefont
  {Schutz}(2009)}]{Sathyaprakash_2009}%
  \BibitemOpen
  \bibfield  {author} {\bibinfo {author} {\bibfnamefont {B.~S.}\ \bibnamefont
  {Sathyaprakash}}\ and\ \bibinfo {author} {\bibfnamefont {B.~F.}\ \bibnamefont
  {Schutz}},\ }\bibfield  {title} {\bibinfo {title} {Physics, astrophysics and
  cosmology with gravitational waves},\ }\bibfield  {journal} {\bibinfo
  {journal} {Living Reviews in Relativity}\ }\textbf {\bibinfo {volume} {12}},\
  \href {https://doi.org/10.12942/lrr-2009-2} {10.12942/lrr-2009-2} (\bibinfo
  {year} {2009}),\ \Eprint {https://arxiv.org/abs/0903.0338} {arXiv:0903.0338
  [gr-qc]} \BibitemShut {NoStop}%
\bibitem [{\citenamefont {Coe}(2009)}]{Coe_2009}%
  \BibitemOpen
  \bibfield  {author} {\bibinfo {author} {\bibfnamefont {D.}~\bibnamefont
  {Coe}},\ }\href@noop {} {\bibinfo {title} {Fisher matrices and confidence
  ellipses: A quick-start guide and software}} (\bibinfo {year} {2009}),\
  \Eprint {https://arxiv.org/abs/0906.4123} {arXiv:0906.4123 [astro-ph.IM]}
  \BibitemShut {NoStop}%
\bibitem [{\citenamefont {Biswas}\ \emph {et~al.}(2020)\citenamefont {Biswas},
  \citenamefont {McIver},\ and\ \citenamefont {Mahabal}}]{Biswas_2020}%
  \BibitemOpen
  \bibfield  {author} {\bibinfo {author} {\bibfnamefont {A.}~\bibnamefont
  {Biswas}}, \bibinfo {author} {\bibfnamefont {J.}~\bibnamefont {McIver}},\
  and\ \bibinfo {author} {\bibfnamefont {A.}~\bibnamefont {Mahabal}},\
  }\bibfield  {title} {\bibinfo {title} {New methods to assess and improve ligo
  detector duty cycle},\ }\href {https://doi.org/10.1088/1361-6382/ab8650}
  {\bibfield  {journal} {\bibinfo  {journal} {Classical and Quantum Gravity}\
  }\textbf {\bibinfo {volume} {37}},\ \bibinfo {pages} {175008} (\bibinfo
  {year} {2020})},\ \Eprint {https://arxiv.org/abs/1910.12143}
  {arXiv:1910.12143 [astro-ph.IM]} \BibitemShut {NoStop}%
\bibitem [{\citenamefont {Chatterji}\ \emph {et~al.}(2004)\citenamefont
  {Chatterji} \emph {et~al.}}]{Chatterji_2004}%
  \BibitemOpen
  \bibfield  {author} {\bibinfo {author} {\bibfnamefont {S.}~\bibnamefont
  {Chatterji}} \emph {et~al.},\ }\bibfield  {title} {\bibinfo {title}
  {Multiresolution techniques for the detection of gravitational-wave bursts},\
  }\href {https://doi.org/10.1088/0264-9381/21/20/024} {\bibfield  {journal}
  {\bibinfo  {journal} {Classical and Quantum Gravity}\ }\textbf {\bibinfo
  {volume} {21}},\ \bibinfo {pages} {S1809} (\bibinfo {year} {2004})},\ \Eprint
  {https://arxiv.org/abs/gr-qc/0412119} {arXiv:gr-qc/0412119 [gr-qc]}
  \BibitemShut {NoStop}%
\bibitem [{\citenamefont {Abbott}\ \emph
  {et~al.}(2016{\natexlab{c}})\citenamefont {Abbott} \emph
  {et~al.}}]{Abbott_2016}%
  \BibitemOpen
  \bibfield  {author} {\bibinfo {author} {\bibfnamefont {B.~P.}\ \bibnamefont
  {Abbott}} \emph {et~al.} (\bibinfo {collaboration} {The LIGO Scientific
  Collaboration and Virgo Collaboration}),\ }\bibfield  {title} {\bibinfo
  {title} {Observation of gravitational waves from a binary black hole
  merger},\ }\bibfield  {journal} {\bibinfo  {journal} {Physical Review
  Letters}\ }\textbf {\bibinfo {volume} {116}},\ \href
  {https://doi.org/10.1103/physrevlett.116.061102}
  {10.1103/physrevlett.116.061102} (\bibinfo {year} {2016}{\natexlab{c}}),\
  \Eprint {https://arxiv.org/abs/1602.03837} {arXiv:1602.03837 [gr-qc]}
  \BibitemShut {NoStop}%
\bibitem [{\citenamefont {Regimbau}\ \emph {et~al.}(2012)\citenamefont
  {Regimbau} \emph {et~al.}}]{Regimbau_2012}%
  \BibitemOpen
  \bibfield  {author} {\bibinfo {author} {\bibfnamefont {T.}~\bibnamefont
  {Regimbau}} \emph {et~al.},\ }\bibfield  {title} {\bibinfo {title} {Mock data
  challenge for the einstein gravitational-wave telescope},\ }\bibfield
  {journal} {\bibinfo  {journal} {Physical Review D}\ }\textbf {\bibinfo
  {volume} {86}},\ \href {https://doi.org/10.1103/physrevd.86.122001}
  {10.1103/physrevd.86.122001} (\bibinfo {year} {2012}),\ \Eprint
  {https://arxiv.org/abs/1201.3563} {arXiv:1201.3563 [gr-qc]} \BibitemShut
  {NoStop}%
\bibitem [{\citenamefont {Bosi}\ and\ \citenamefont
  {Porter}(2011)}]{Bosi_2010}%
  \BibitemOpen
  \bibfield  {author} {\bibinfo {author} {\bibfnamefont {L.}~\bibnamefont
  {Bosi}}\ and\ \bibinfo {author} {\bibfnamefont {E.~K.}\ \bibnamefont
  {Porter}},\ }\bibfield  {title} {\bibinfo {title} {Data analysis challenges
  for the einstein telescope},\ }\href
  {https://doi.org/10.1007/s10714-010-1084-3} {\bibfield  {journal} {\bibinfo
  {journal} {General Relativity and Gravitation}\ }\textbf {\bibinfo {volume}
  {43}},\ \bibinfo {pages} {519} (\bibinfo {year} {2011})},\ \Eprint
  {https://arxiv.org/abs/0910.0380} {arXiv:0910.0380 [gr-qc]} \BibitemShut
  {NoStop}%
\bibitem [{\citenamefont {Punturo}\ \emph {et~al.}(2010)\citenamefont {Punturo}
  \emph {et~al.}}]{Punturo_2010}%
  \BibitemOpen
  \bibfield  {author} {\bibinfo {author} {\bibfnamefont {M.}~\bibnamefont
  {Punturo}} \emph {et~al.},\ }\bibfield  {title} {\bibinfo {title} {The
  einstein telescope: A third-generation gravitational wave observatory},\
  }\href {https://doi.org/10.1088/0264-9381/27/19/194002} {\bibfield  {journal}
  {\bibinfo  {journal} {Classical and Quantum Gravity}\ }\textbf {\bibinfo
  {volume} {27}},\ \bibinfo {pages} {194002} (\bibinfo {year}
  {2010})}\BibitemShut {NoStop}%
\bibitem [{\citenamefont {Abbott}\ \emph
  {et~al.}(2020{\natexlab{c}})\citenamefont {Abbott} \emph
  {et~al.}}]{Abbott_2020_GW190814}%
  \BibitemOpen
  \bibfield  {author} {\bibinfo {author} {\bibfnamefont {R.}~\bibnamefont
  {Abbott}} \emph {et~al.} (\bibinfo {collaboration} {The LIGO Scientific
  Collaboration and Virgo Collaboration}),\ }\bibfield  {title} {\bibinfo
  {title} {Gw190814: Gravitational waves from the coalescence of a 23 solar
  mass black hole with a 2.6 solar mass compact object},\ }\href
  {https://doi.org/10.3847/2041-8213/ab960f} {\bibfield  {journal} {\bibinfo
  {journal} {The Astrophysical Journal}\ }\textbf {\bibinfo {volume} {896}},\
  \bibinfo {pages} {L44} (\bibinfo {year} {2020}{\natexlab{c}})},\ \Eprint
  {https://arxiv.org/abs/2006.12611} {arXiv:2006.12611 [astro-ph.HE]}
  \BibitemShut {NoStop}%
\bibitem [{\citenamefont {Berry}\ \emph {et~al.}(2015)\citenamefont {Berry}
  \emph {et~al.}}]{Berry_2015}%
  \BibitemOpen
  \bibfield  {author} {\bibinfo {author} {\bibfnamefont {C.~P.~L.}\
  \bibnamefont {Berry}} \emph {et~al.},\ }\bibfield  {title} {\bibinfo {title}
  {Parameter estimation for binary neutron-star coalescences with realistic
  noise during the advanced ligo era},\ }\href
  {https://doi.org/10.1088/0004-637x/804/2/114} {\bibfield  {journal} {\bibinfo
   {journal} {The Astrophysical Journal}\ }\textbf {\bibinfo {volume} {804}},\
  \bibinfo {pages} {114} (\bibinfo {year} {2015})},\ \Eprint
  {https://arxiv.org/abs/1411.6934} {arXiv:1411.6934 [astro-ph.HE]}
  \BibitemShut {NoStop}%
\bibitem [{\citenamefont {Pozzo}(2014)}]{Del_Pozzo_2014}%
  \BibitemOpen
  \bibfield  {author} {\bibinfo {author} {\bibfnamefont {W.~D.}\ \bibnamefont
  {Pozzo}},\ }\bibfield  {title} {\bibinfo {title} {Measuring the hubble
  constant using gravitational waves},\ }\href
  {https://doi.org/10.1088/1742-6596/484/1/012030} {\bibfield  {journal}
  {\bibinfo  {journal} {Journal of Physics: Conference Series}\ }\textbf
  {\bibinfo {volume} {484}},\ \bibinfo {pages} {012030} (\bibinfo {year}
  {2014})}\BibitemShut {NoStop}%
\end{thebibliography}%

\end{document}